\RequirePackage{fix-cm}

\documentclass[twocolumn,epjc3]{svjour3}  
\smartqed  
\RequirePackage{mathptmx}      
\RequirePackage{graphicx}
\RequirePackage{amssymb}
\RequirePackage{amsmath}
\RequirePackage{lineno}
\RequirePackage{booktabs}
\RequirePackage{float}
\RequirePackage{comment}
\RequirePackage{mhchem}
\RequirePackage{booktabs}
\RequirePackage{url}
\RequirePackage[colorlinks,allcolors=red]{hyperref}
\RequirePackage[multiple]{footmisc}
\RequirePackage{upgreek}
\RequirePackage{mathtools}
\RequirePackage{cuted}
\RequirePackage{relsize}
\RequirePackage{makecell}
\RequirePackage{tablefootnote}
\RequirePackage{siunitx}
\RequirePackage{relsize}
\RequirePackage[english]{babel}
\RequirePackage{xspace}
\RequirePackage{multirow}

\newcommand{\Alberta}{Department of Physics, University of Alberta, Edmonton, AB T6G 2R3, Canada}
\newcommand{\APC}{APC, Universit\'e de Paris Cit\'e, CNRS, Astroparticule et Cosmologie, Paris F-75013, France}
\newcommand{\AQLNGS}{INFN Laboratori Nazionali del Gran Sasso, Assergi (AQ) 67100, Italy}
\newcommand{\AQGSSI}{Gran Sasso Science Institute, L'Aquila 67100, Italy}

\newcommand{\AstroCeNT}{AstroCeNT, Nicolaus Copernicus Astronomical Center of the Polish Academy of Sciences, 00-614 Warsaw, Poland}
\newcommand{\Augustana}{Physics Department, Augustana University, Sioux Falls, SD 57197, USA}
\newcommand{\Belgorod}{Radiation Physics Laboratory, Belgorod National Research University, Belgorod 308007, Russia}

\newcommand{\BINP}{Budker Institute of Nuclear Physics, Novosibirsk 630090, Russia}
\newcommand{\Birmingham}{School of Physics and Astronomy, University of Birmingham, Edgbaston, B15 2TT, Birmingham, UK}

\newcommand{\BOINFN}{INFN Bologna, Bologna 40126, Italy}
\newcommand{\BOUniPHY}{Department of Physics and Astronomy, Universit\`a degli Studi di Bologna, Bologna 40126, Italy}
\newcommand{\CAUniCHE}{Department of Mechanical, Chemical, and Materials Engineering, Universit\`a degli Studi, Cagliari 09042, Italy}
\newcommand{\CAUniEEE}{Department of Electrical and Electronic Engineering, Universit\`a degli Studi di Cagliari, Cagliari 09123, Italy}
\newcommand{\CAUniPHY}{Physics Department, Universit\`a degli Studi di Cagliari, Cagliari 09042, Italy}
\newcommand{\CAINFN}{INFN Cagliari, Cagliari 09042, Italy}
\newcommand{\Carleton}{Department of Physics, Carleton University, Ottawa, ON K1S 5B6, Canada}

\newcommand{\Columbia}{Physics Department, Columbia University, New York, NY 10027, USA}
\newcommand{\Chicago}{Department of Physics and Kavli Institute for Cosmological Physics, University of Chicago, Chicago, IL 60637, USA}
\newcommand{\BOCentroFermi}{Museo Storico della Fisica e Centro Studi e Ricerche Enrico Fermi, Roma 00184, Italy}

\newcommand{\CIEMAT}{CIEMAT, Centro de Investigaciones Energ\'eticas, Medioambientales y Tecnol\'ogicas, Madrid 28040, Spain}

\newcommand{\CPPM}{Centre de Physique des Particules de Marseille, Aix Marseille Univ, CNRS/IN2P3, CPPM, Marseille, France}
\newcommand{\CTINFN}{INFN Catania, Catania 95121, Italy}
\newcommand{\CTUNI}{Universit\`a of Catania, Catania 95124, Italy}
\newcommand{\CTLNS}{INFN Laboratori Nazionali del Sud, Catania 95123, Italy}

\newcommand{\ENUniCEE}{Engineering and Architecture Faculty, Universit\`a di Enna Kore, Enna 94100, Italy}
\newcommand{\ETHZ}{Institute for Particle Physics, ETH Z\"urich, Z\"urich 8093, Switzerland}

\newcommand{\FortLewis}{Department of Physics and Engineering, Fort Lewis College, Durango, CO 81301, USA}
\newcommand{\GEUni}{Physics Department, Universit\`a degli Studi di Genova, Genova 16146, Italy}
\newcommand{\GEINFN}{INFN Genova, Genova 16146, Italy}
\newcommand{\Hawaii}{Department of Physics and Astronomy, University of Hawai'i, Honolulu, HI 96822, USA}
\newcommand{\Houston}{Department of Physics, University of Houston, Houston, TX 77204, USA}
\newcommand{\IHEP}{Institute of High Energy Physics, Chinese Academy of Sciences, Beijing 100049, China}
\newcommand{\INFN}{Istituto Nazionale di Fisica Nucleare, Roma 00186, Italia}

\newcommand{\JINR}{Joint Institute for Nuclear Research, Dubna 141980, Russia}
\newcommand{\Krakow}{M.~Smoluchowski Institute of Physics, Jagiellonian University, 30-348 Krakow, Poland}
\newcommand{\Kurchatov}{National Research Centre Kurchatov Institute, Moscow 123182, Russia}
\newcommand{\Laurentian}{Department of Physics and Astronomy, Laurentian University, Sudbury, ON P3E 2C6, Canada}
\newcommand{\Lancaster}{Physics Department, Lancaster University, Lancaster LA1 4YB, UK}
\newcommand{\LFoundry}{LFoundry S.r.l, Via Antonio Pacinotti 7, Avezzano AQ 67051, IT}
\newcommand{\Liverpool}{Department of Physics, University of Liverpool, The Oliver Lodge Laboratory, Liverpool L69 7ZE, UK}

\newcommand{\LNLINFN}{INFN Laboratori Nazionali di Legnaro, Legnaro (Padova) 35020, Italy}
\newcommand{\Lodz}{Institute of Applied Radiation Chemistry, Lodz University of Technology, 93-590 Lodz, Poland}

\newcommand{\Manchester}{Department of Physics and Astronomy, The University of Manchester, Manchester M13 9PL, UK}
\newcommand{\MEPhI}{National Research Nuclear University MEPhI, Moscow 115409, Russia}
\newcommand{\MendeleevUniverisity}{Mendeleev University of Chemical Technology, Moscow 125047, Russia}

\newcommand{\MIINFN}{INFN Milano, Milano 20133, Italy}
\newcommand{\MIPoliICA}{Civil and Environmental Engineering Department, Politecnico di Milano, Milano 20133, Italy}
\newcommand{\MIPoliCHE}{Chemistry, Materials and Chemical Engineering Department ``G.~Natta", Politecnico di Milano, Milano 20133, Italy}

\newcommand{\MIUni}{Physics Department, Universit\`a degli Studi di Milano, Milano 20133, Italy}
\newcommand{\MSU}{Skobeltsyn Institute of Nuclear Physics, Lomonosov Moscow State University, Moscow 119234, Russia}
\newcommand{\NAINFN}{INFN Napoli, Napoli 80126, Italy}
\newcommand{\NAUniPHY}{Physics Department, Universit\`a degli Studi ``Federico II'' di Napoli, Napoli 80126, Italy}
\newcommand{\NAUniCHE}{Chemical, Materials, and Industrial Production Engineering Department, Universit\`a degli Studi ``Federico II'' di Napoli, Napoli 80126, Italy}
\newcommand{\NAUniDIST}{Department of Structures for Engineering and Architecture, Universit\`a degli Studi ``Federico II'' di Napoli, Napoli 80126, Italy}
\newcommand{\NAUniPHARM}{Pharmacy Department, Universit\`a degli Studi ``Federico II'' di Napoli, Napoli 80131, Italy}

\newcommand{\Oxford}{University of Oxford, Oxford OX1 2JD, United Kingdom}
\newcommand{\Petersburg}{Saint Petersburg Nuclear Physics Institute, Gatchina 188350, Russia}

\newcommand{\PIINFN}{INFN Pisa, Pisa 56127, Italy}
\newcommand{\PIUniPHY}{Physics Department, Universit\`a degli Studi di Pisa, Pisa 56127, Italy}
\newcommand{\PNNL}{Pacific Northwest National Laboratory, Richland, WA 99352, USA}
\newcommand{\Princeton}{Physics Department, Princeton University, Princeton, NJ 08544, USA}
\newcommand{\Queens}{Department of Physics, Engineering Physics and Astronomy, Queen's University, Kingston, ON K7L 3N6, Canada}
\newcommand{\RHUL}{Department of Physics, Royal Holloway University of London, Egham TW20 0EX, UK}
\newcommand{\RMTreINFN}{INFN Roma Tre, Roma 00146, Italy}

\newcommand{\RMUnoINFN}{INFN Sezione di Roma, Roma 00185, Italy}
\newcommand{\RMUnoUni}{Physics Department, Sapienza Universit\`a di Roma, Roma 00185, Italy}

\newcommand{\SNL}{Savannah River National Laboratory, Jackson, SC 29831, United States}

\newcommand{\SNOLAB}{SNOLAB, Lively, ON P3Y 1N2, Canada}

\newcommand{\STFCInterconnect}{Science \& Technology Facilities Council (STFC), Rutherford Appleton Laboratory, Technology, Harwell Oxford, Didcot OX11 0QX, UK}

\newcommand{\Temple}{Physics Department, Temple University, Philadelphia, PA 19122, USA}
\newcommand{\TNFBK}{Fondazione Bruno Kessler, Povo 38123, Italy}

\newcommand{\TOINFN}{INFN Torino, Torino 10125, Italy}
\newcommand{\TOPoli}{Department of Electronics and Telecommunications, Politecnico di Torino, Torino 10129, Italy}

\newcommand{\TRIUMF}{TRIUMF, 4004 Wesbrook Mall, Vancouver, BC V6T 2A3, Canada}

\newcommand{\UCDavis}{Department of Physics, University of California, Davis, CA 95616, USA}
\newcommand{\UCRiverside}{Department of Physics and Astronomy, University of California, Riverside, CA 92507, USA}

\newcommand{\UCLA}{Physics and Astronomy Department, University of California, Los Angeles, CA 90095, USA}
\newcommand{\UCAS}{University of Chinese Academy of Sciences, Beijing 100049, China}
\newcommand{\UMass}{Amherst Center for Fundamental Interactions and Physics Department, University of Massachusetts, Amherst, MA 01003, USA}

\newcommand{\UnivAQ}{Universit\`a degli Studi dell'Aquila, L'Aquila 67100, Italy}
\newcommand{\UniversityofEdinburgh}{School of Physics and Astronomy, University of Edinburgh, Edinburgh EH9 3FD, UK}

\newcommand{\USP}{Instituto de F\'isica, Universidade de S\~ao Paulo, S\~ao Paulo 05508-090, Brazil}
\newcommand{\VTech}{Virginia Tech, Blacksburg, VA 24061, USA}
\newcommand{\Warwick}{University of Warwick, Department of Physics, Coventry CV47AL, UK}
\newcommand{\WUT}{Institute of Radioelectronics and Multimedia Technology, Warsaw University of Technology, 00-661 Warsaw, Poland}
\newcommand{\WilliamsCollege}{Williams College, Physics Department, Williamstown, MA 01267 USA}
\newcommand{\Zaragoza}{Centro de Astropart\'iculas y F\'isica de Altas Energ\'ias, Universidad de Zaragoza, Zaragoza 50009, Spain}

\newcommand{\UniHAM}{Institute of Experimental Physics, University of Hamburg, Luruper Chaussee 149, 22761, Hamburg, Germany}
\newcommand{\Seattle}{Center for Experimental Nuclear Physics and Astrophysics, and Department of Physics, University of Washington, Seattle, WA 98195, USA} 

\newcommand{\RomanNumeralCaps}[1]
    {\MakeUppercase{\romannumeral #1}}

\newcommand{\NWAFER}{1314\xspace}
\newcommand{\PERCENTPRODUCTION}{94}

\newcommand{\NSIPM}{359040\xspace}
\newcommand{\NSIPMVBDout}{2698\xspace}
\newcommand{\NSIPMRQout}{2517\xspace}
\newcommand{\NSIPMGOFout}{9261\xspace}
\newcommand{\NSIPMIout}{2403\xspace}

\newcommand{\LOT}{9306869\xspace}
\newcommand{\WAFER}{19\xspace}

\journalname{Eur. Phys. J. C}

\makeatletter
\renewcommand{\thanksref}[1]{\nolinebreak\textsuperscript{\ref{#1}}\nolinebreak\checknextarg}
\newcommand{\checknextarg}{\@ifnextchar\bgroup{\nolinebreak\gobblenextarg}{}}
\newcommand{\gobblenextarg}[1]{ \textsuperscript{\nolinebreak\hspace{-4pt}\mbox{\nolinebreak$^,$\nolinebreak\ref{#1}\nolinebreak}\nolinebreak} \@ifnextchar\bgroup{\gobblenextarg}{}}

\begin{document}

\title{Quality Assurance and Quality Control of the $26~\text{m}^2$ SiPM production for the DarkSide-20k dark matter experiment}
 \widowpenalty10000
  \clubpenalty10000

\author{The DarkSide-20k Collaboration$^\text{\normalfont a,1}$}
\thankstext{e1}{\vspace{1cm}*corresponding authors: ds-ed@lists.infn.it}
\institute{See back for author list \label{addr1}}

\date{Received: date / Accepted: date}

\maketitle

\modulolinenumbers[5]

\begin{abstract}
DarkSide-20k is a novel liquid argon dark matter detector currently under construction at the Laboratori Nazionali del Gran Sasso (LNGS) of the Istituto Nazionale di Fisica Nucleare (INFN) that will push the sensitivity for Weakly Interacting Massive Particle (WIMP) detection into the neutrino fog. The core of the apparatus is a dual-phase Time Projection Chamber (TPC), filled with \SI{50} {tonnes} of low radioactivity underground argon (UAr) acting as the WIMP target. 
NUV-HD-cryo Silicon Photomultipliers (SiPM)s designed by Fondazione Bruno Kessler (FBK) (Trento, Italy) were selected as the photon sensors covering two $10.5~\text{m}^2$ Optical Planes, one at each end of the TPC, and a total of $5~\text{m}^2$ photosensitive surface for the liquid argon veto detectors. This paper describes the Quality Assurance and Quality Control (QA/QC) plan and procedures accompanying the production of FBK~NUV-HD-cryo SiPM wafers manufactured by LFoundry s.r.l. (Avezzano, AQ, Italy).
SiPM characteristics are measured at 77~K at the wafer level with a custom-designed probe station. As of March~2025, \NWAFER of the 1400  production wafers (\PERCENTPRODUCTION\% of the total) for DarkSide-20k were tested. The wafer yield is $93.2\pm2.5$\%, which exceeds the 80\% specification defined in the original DarkSide-20k production plan. 
\end{abstract}

\section{Introduction}
\label{sec:introduction}

Silicon Photomultipliers (SiPMs) have emerged as a compelling photosensor solution for detecting single photons in applications ranging from particle physics to medical imaging and beyond~\cite{Acerbi2018}. SiPMs consist of an array of tightly packaged Single Photon Avalanche Diodes (SPADs) operated above the breakdown voltage, $V_\mathrm{bd}$, so that they generate self sustaining charge avalanches upon absorbing an incident photon~\cite{Gallina2019avalanche}.  In contrast to the widely used Photo-multiplier Tubes (PMTs), SiPMs are low-voltage powered, insensitive to magnetic field, and have a compact and flat form factor~\cite{Baudis2018}. For these reasons, SiPMs are the adopted solution in the MEG-\RomanNumeralCaps{2} and DUNE experiments~\cite{Baldini2018,FALCONE2021164648}. In addition, SiPMs have very low residual natural radioactivity, making them especially appealing for low-background experiments such as nEXO and DarkSide-20k~\cite{Ako,Aalseth2018}.

DarkSide-20k (DS-20k) is a multi-tonne dark matter detector under construction at the Laboratori Nazionali del Gran Sasso (LNGS) of the Istituto Nazionale di Fisica Nucleare (INFN) that will push the sensitivity for Weakly Interacting Massive Particles (WIMP) detection to the level where solar and atmospheric neutrinos become a significant background.  The core of the DS-20k apparatus is a dual-phase Time Projection Chamber (TPC), with a vertical electron drift. The TPC is a \SI{348}{cm} tall octagonal prism with a \SI{350}{cm} inner diameter containing the liquid argon (LAr) dark matter target: \SI{50} {tonnes} of underground argon (UAr) (\SI{20}{t} fiducial volume)~\cite{Aaron_2023}. 

Electron extraction in gas is provided by a stainless steel grid made with 200~$\mu\text{m}$ wires, with a pitch of \SI{3}{mm}. The gas pocket thickness is expected to be \SI{7}{mm}. The top and bottom lids of the octagon are made in pure acrylic, and their inner planes are coated with a thin conductive layer of PEDOT:PSS (Clevios\textsuperscript{\texttrademark}) and tetraphenyl butadiene (TPB) to shift  LAr scintillation light from \SI{128}{nm} to visible light~\cite{Benson2018}. The conductive layers allow biasing the inner planes of the two lids,
such that they act as cathode and anode.  The barrel pieces of the octagon are machined and coated with Clevios\textsuperscript{\texttrademark} to define a geometry of conductive rings (field cage) connected by resistors and biased to define a uniform and stable electric drift field of \SI{200}{V/cm}.

The scintillation light emitted in the TPC is detected by two $10.5~\text{m}^2$ Optical Planes (OP) instrumented with Fondazione Bruno Kessler (FBK) NUV-HD-cryo (near-UV sensitive, high-density, cryogenic compatible) SiPMs located at the top and bottom of the octagonal barrel and used to optically readout the argon scintillation (S1) and electroluminescence (S2) signals in the liquid and gas phase, respectively~\cite{Agnes:2015gu,Agnes:2018ep}. 

A stainless steel vessel filled with an additional \SI{36} {tonnes} of UAr surrounds the TPC. The stainless steel vessel is itself immersed in further \SI{650}{tonnes} of atmospheric liquid argon (AAr), contained within a DUNE-like membrane cryostat. Both argon volumes are outfitted with SiPMs and will serve as the experiment's Inner Veto (IV) and Outer Veto (OV) detectors~\cite{Santone:2024sc}. 

The inner detector employs the TPC and IV detectors to mitigate the most critical background for dark matter searches, which comes from neutron scattering, while the OV is used to detect and tag external neutrons and muons.

The FBK NUV-HD SiPM technology was introduced in 2016~\cite{piemonte2016performance}. The first generation of these devices 
suffered from a  Dark Count Rate (DCR) and afterpulsing probability higher than DS-20k specifications at cryogenic temperatures~\cite{Aalseth2018}. A specific process modification was developed by FBK to address the DS-20k requirements~\cite{Acerbi2017}. This led to a variant of the FBK NUV-HD technology named FBK NUV-HD-LF (Low Field) that features an almost three-orders of magnitude reduction of the SiPM DCR at 80~K when compared to the standard NUV-HD version, comfortably lower than the DS-20k specification.

\begin{figure}[ht]
\centering
\centering\includegraphics[width=0.99\linewidth]{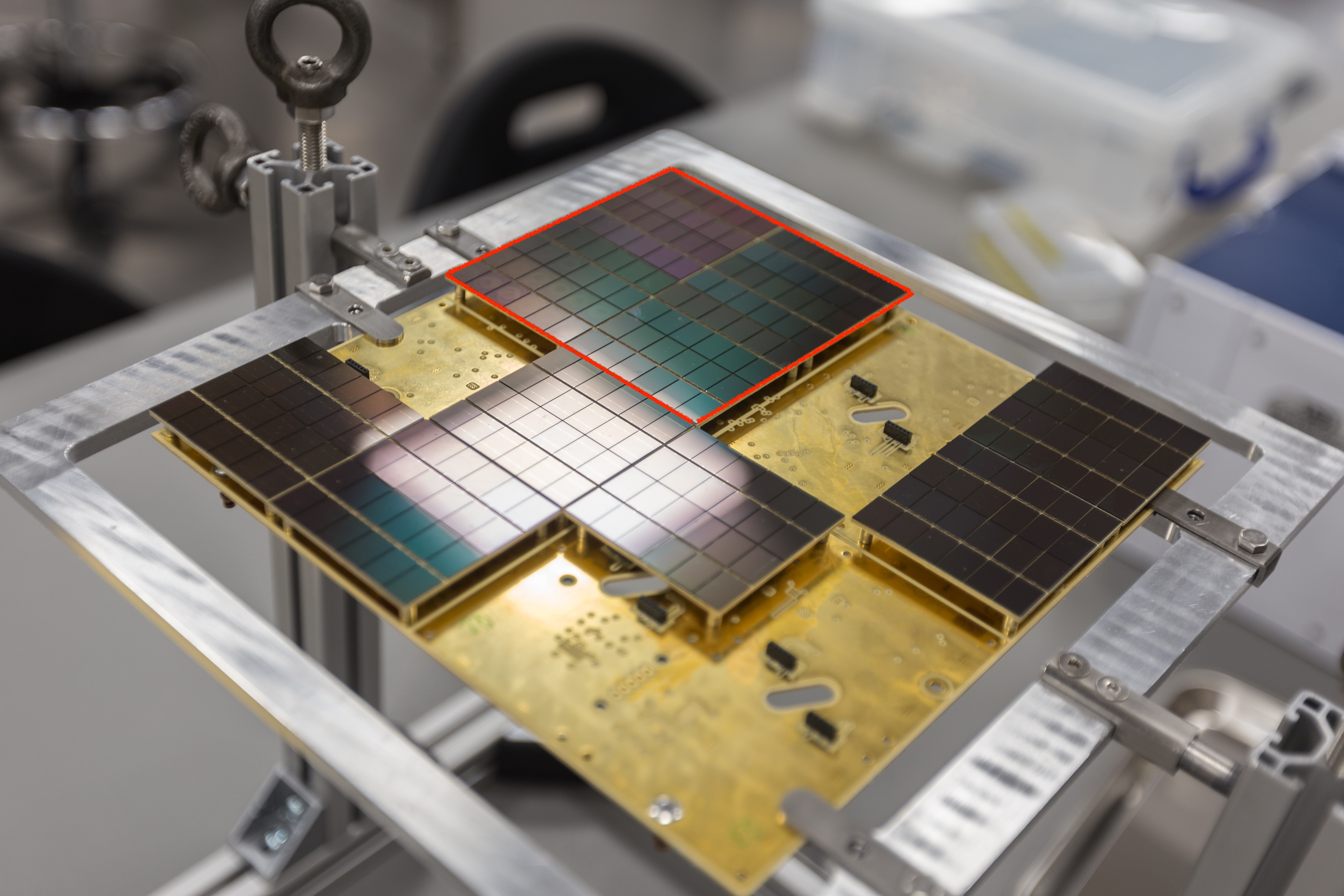}
\caption{Tiles prototype partially assembled into a Photo Detector Units (PDU). Each tile consists of 24 FBK NUV-HD-cryo SiPMs with six (three) parallel branches of four (two) SiPMs in series fed into a single transimpedance amplifier (TIA) or a custom designed ASIC amplifier. Tiles, in groups of four, are further aggregated in quadrants. Only one quadrant, shown in red in the figure, is fully populated. Each quadrant is read out as an individual analog readout channel.}
\label{fig:tiles}
\end{figure}

Further optimization of this technology  allowed for a reduced SiPM afterpulsing probability (NUV-HD-lowAP technology) and quenching resistance, thus optimizing the microcell recharge time constant at cryogenic temperature. These two process adjustments led to the development of a new technology called FBK NUV-HD-cryo, presented in Ref.~\cite{Gola2019}, that not only met but exceeded the DS-20k specifications.\\In the final DS-20k photodetector design 24 FBK NUV-HD-cryo SiPMs are aggregated together to form objects known as tiles, with six (three) parallel branches of four (two) SiPMs in series fed into a single transimpedance amplifier (TIA) or a custom designed ASIC amplifier~\cite{DIncecco2018,Kugathasan:2020xry}, used to instrument the TPC and Veto detectors~\cite{Franchini_2024}. The 24 SiPMs are bonded to an Arlon low-radioactive Printed Circuit Board (PCB)~\cite{Arlon_doc}.\\Tiles, in groups of four, are further aggregated in a quadrant: a single analog readout element~\cite{Razeto2022}. Four quadrants are then aggregated into a single object with a total surface area of $400~\text{cm}^2$. These devices are called Photo Detector Units (PDUs) and constitute the basic element of the DS-20k photoelectronic system. An example is shown in Fig.~\ref{fig:tiles}. In total, 680 PDUs are needed to instrument the two TPC optical planes and the IV and OV detectors (528 TPC, 152 IV and OV), which will make DS-20k the largest SiPM-based cryogenic detector, with more than 260,000 SiPMs.  

This paper focuses on the Quality Assurance and Quality Control (QA/QC) of the FBK~NUV-HD-cryo SiPM production wafers manufactured by LFoundry s.r.l. (Avezzano, AQ, Italy), an Italian semiconductor company with the infrastructure to produce a large number of devices for DS-20k~\cite{ORGANTINI2020164410}. 

Wafers manufactured by LFoundry are stored, tested and diced in the  Nuova Officina Assergi (NOA) facility, a new ISO-6  $353~\text{m}^2$ clean room packaging facility constructed at INFN LNGS~\cite{Consiglio2023,KOCHANEK2020164487,10.3389/fphy.2024.1433347}. NOA houses state-of-the-art production machines for packaging silicon sensors such as the ones used in DS-20k, and it is currently used for the production of the DS-20k PDUs.

Several SiPM characteristics for each SiPM dice such as breakdown voltage, quenching resistance, leakage current and correlated avalanche noise, are measured with a custom-designed, cryogenic probe station at the wafer level at 77~K. As of March~2025, \NWAFER of the 1400 production wafers (\PERCENTPRODUCTION\% of the SiPM production) were tested.  Additionally, 46 extra wafers, considered as engineering pre-production runs, were evaluated. Although identical to those used in the final production, these wafers are termed "engineering" because they were used for prototype tile assembly rather than for fabricating final production-grade tiles. The average wafer yield is $93.2\pm2.5$\%, which exceeds the 80\% production yield assumed in the original DS-20k production plan. 

\section{Hardware Setup}
\label{sec:hardware}

\begin{figure}[ht]
\centering\includegraphics[width=0.99\linewidth]{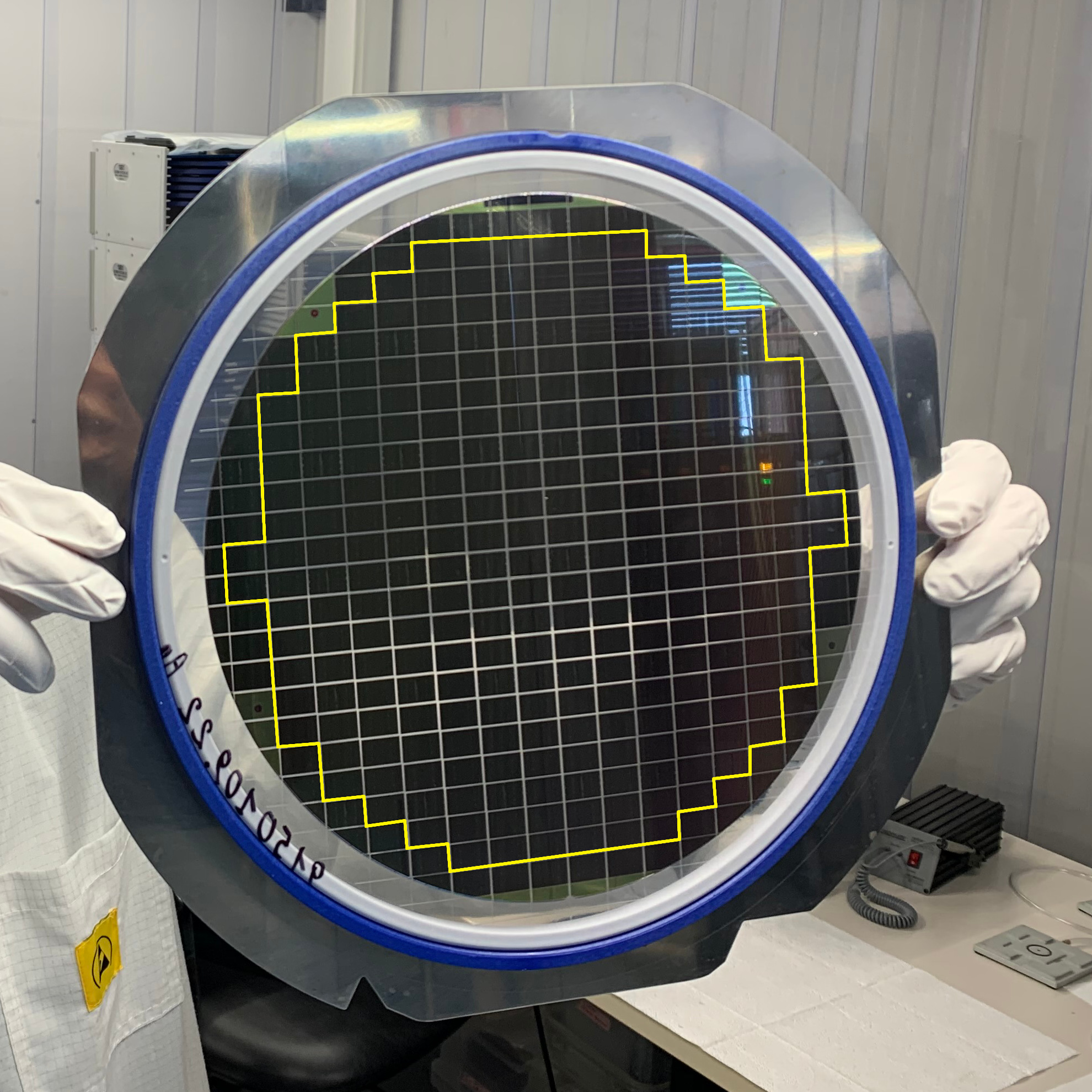}
\caption{Production wafer diced and mounted on a grip-ring. The 268 SiPM dice are highlighted in yellow. Due to the overlapping geometry of the probe card and the wafer clamping ring, only 264 of the 268 potentially working SiPMs can be probed at 77~K.}
\label{fig:wafer}
\end{figure}

Fig.~\ref{fig:wafer} shows an 8-inch FBK NUV-HD-cryo SiPM production wafers manufactured by LFoundry. Each wafer consists of 268 potentially functional dice, highlighted in yellow in the figure, with a dimension of 11.7$\times$\SI{7.9}{m\meter^2}. The wafer in the figure is diced and mounted on a grip-ring, used in the subsequent steps of the DS-20k silicon packaging process flow. 

LFoundry produced wafers in 57 batches, known as Lots. Each Lot, containing approximately 25 wafers, is processed together through every foundry stage using automated handling. The most significant variations in wafer performance are expected when comparing different Lots, whereas wafer-to-wafer variations within a single Lot are anticipated to be a minor factor. For batched processes, the differences between wafers are mainly attributable to their positions within the processing equipment. In contrast, for single-process steps, variability arises primarily from tool performance fluctuations over time\footnote{An additional component of wafer variability is within-wafer, or site-to-site, variation. This occurs due to process non-uniformities at the wafer scale.}.
\begin{figure}[ht]
\centering\includegraphics[width=0.99\linewidth]{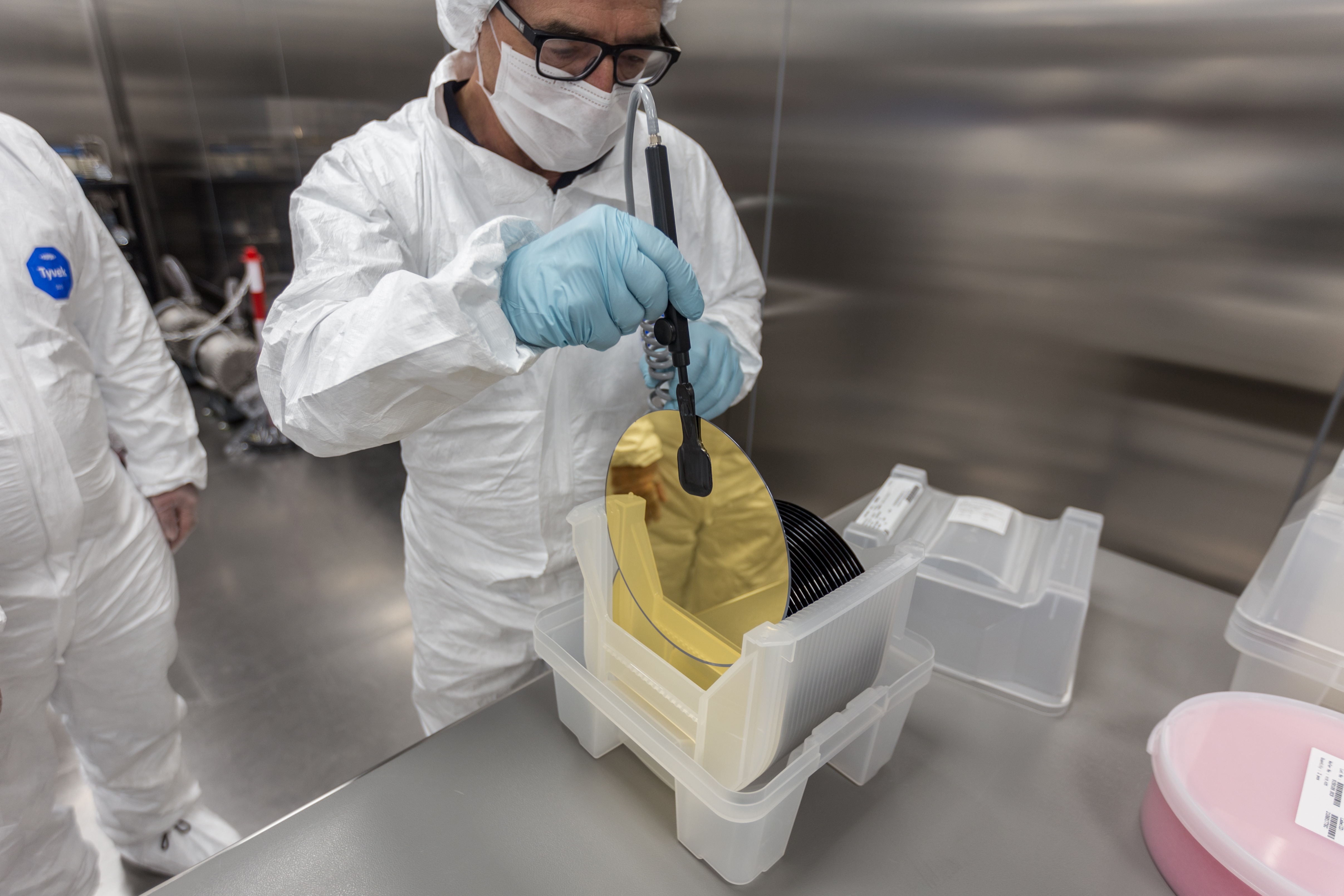}
\caption{Wafer cassette containing one production Lot (25 wafers). The wafer's gold backside is visible.}
\label{fig:wafercassette}
\end{figure}

\begin{figure}[ht]
\centering\includegraphics[width=0.99\linewidth]{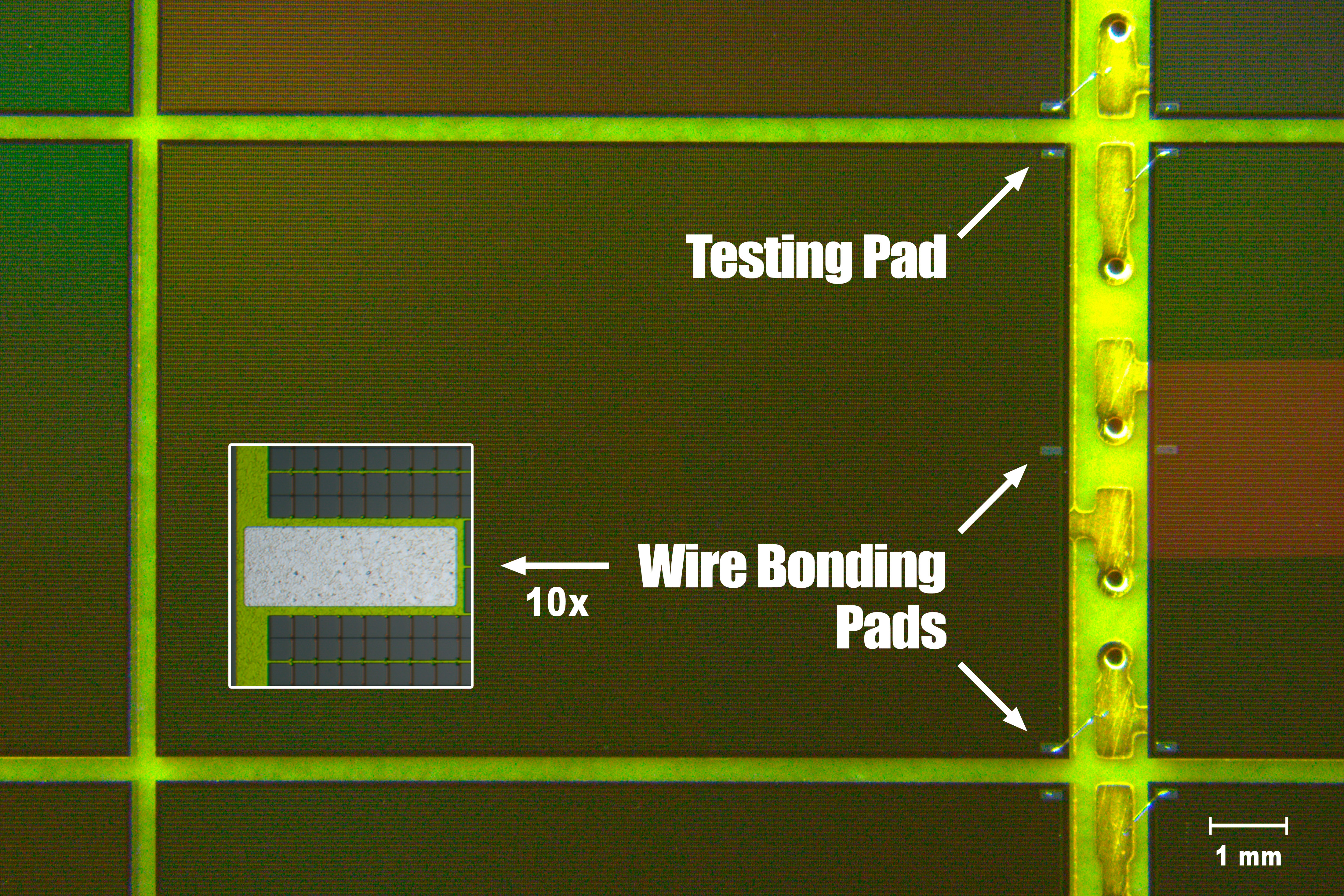}
\caption{FBK NUV-HD-cryo SiPM dice bonded to a DS-20k tile. The three aluminum pads are short-circuited together. One of these three pads is used for cryoprobing (top-right), while the other two are used to wire-bond the SiPMs to the corresponding tile PCB. Each dice measures 11.7$\times$\SI{7.9}{m\meter^2}.}
\label{fig:sipm_pad}
\end{figure}

Fig~\ref{fig:wafercassette} shows one of the production Lots. Each wafer in the Lot has a gold-coated backside ({\it i.e.} silicon substrate) that acts as the SiPM cathode. The SiPM anode contact is made with three bonding pads, connected together, on the frontside of the chips (Fig.~\ref{fig:sipm_pad}). One of these three pads is used for cryoprobing (top-right pad in Fig.~\ref{fig:sipm_pad}), and the other two are used to wire bond the SiPM to the tile PCB.

\begin{figure*}[ht]
\centering
\centering\includegraphics[width=0.8\linewidth]{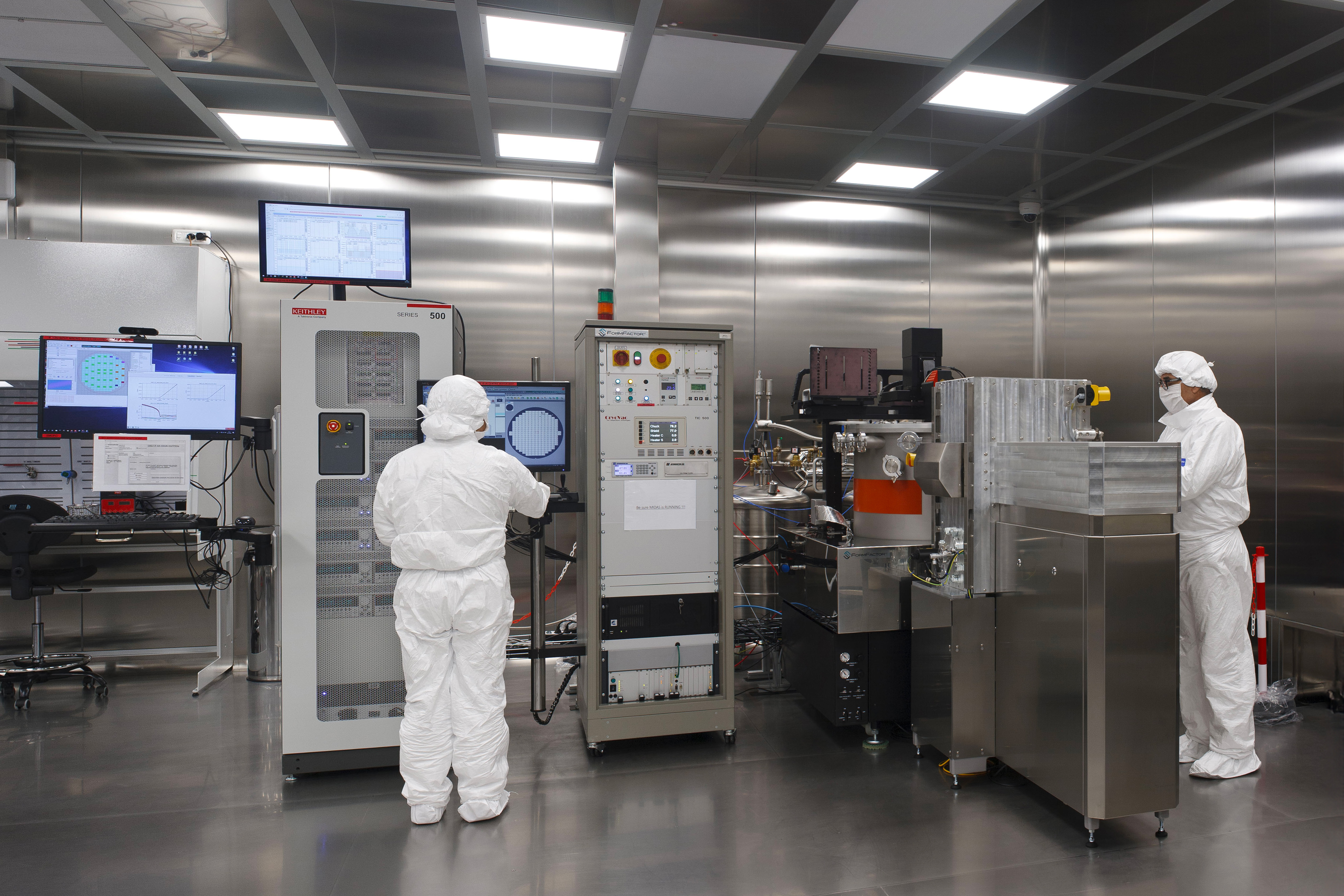}
\caption{FormFactor PAC200 probe system + S500 Parametric Keithley tester used to characterize the response of the FBK NUV-HD-cryo SiPMs. The entire system is located in a new ISO-6  $353~\text{m}^2$ clean room packaging facility called the Nuova Officina Assergi (NOA) constructed at the Istituto Nazionale di Fisica Nucleare (INFN) Laboratori Nazionali del Gran Sasso (LNGS).}
\label{fig:smucp}
\end{figure*}

\begin{figure}[ht]
\centering
\centering\includegraphics[width=0.99\linewidth]{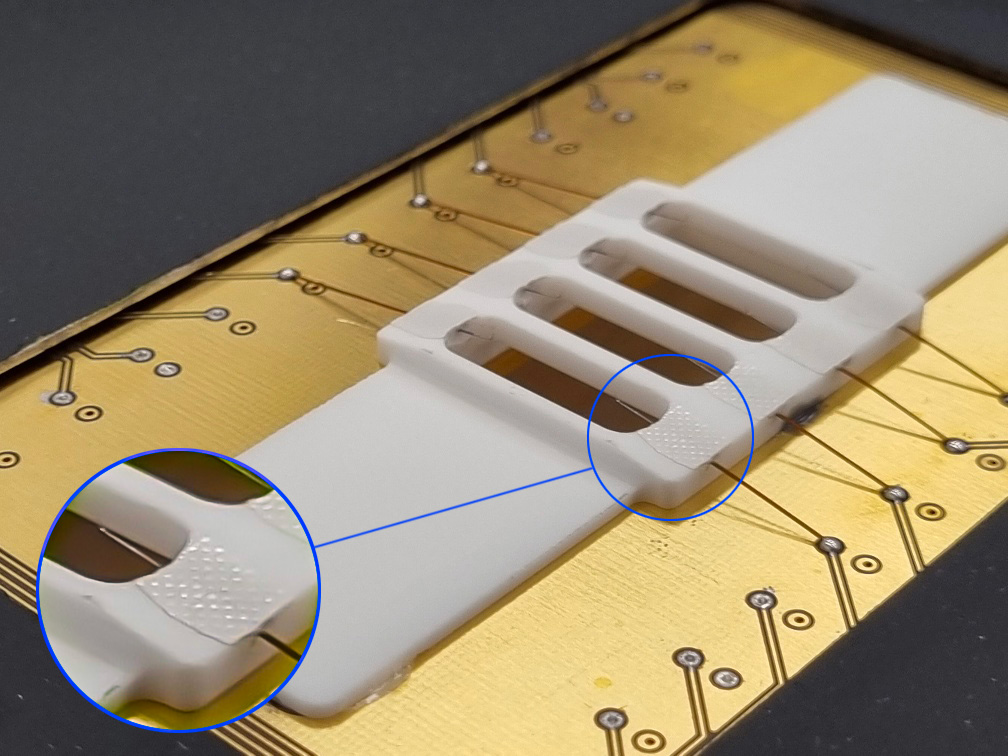}
\caption{Htt probe cards designed by FormFactor with 2x4 tungsten needles used for SiPM probing (1 needle per SiPM - shown enlarged in the circle). The white ceramic carrier shown in this photograph connects directly to the needles.}
\label{fig:htt_probecard}
\end{figure}

A new testing setup (Fig.~\ref{fig:smucp}) was built to characterize the SiPM response at the wafer level at 77~K.
It consists of a high precision, semi-automated cryogenic Cascade FormFactor PAC200 probe system (cryoprobe) that cools the wafer to 77~K in a high vacuum environment ($10^{-7}~\text{mbar}$) \cite{FormFactor}. Wafers are tested with a cantilever, needle-based probe card made by htt group (Munich, Germany)~\cite{htt} (Fig.~\ref{fig:htt_probecard}). 

\begin{figure}[ht]
\centering
\centering\includegraphics[width=0.99\linewidth]{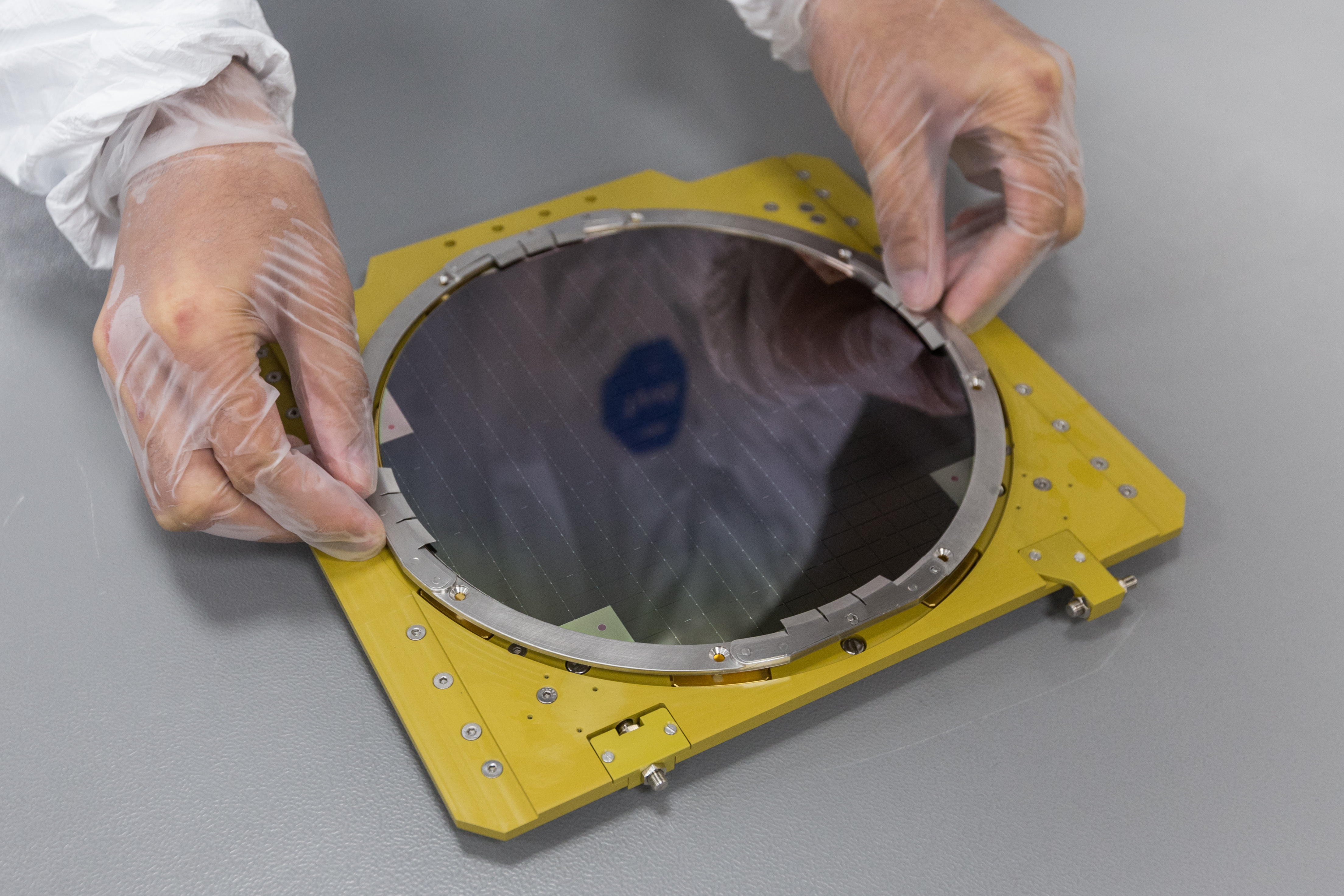}
\caption{Production wafer mounted on a gold-plated wafer carrier. The wafer is held in place with a custom-made clamping ring.}
\label{fig:wafercarrier}
\end{figure}

For the test, wafers are mounted on a gold-plated copper carrier (Fig.~\ref{fig:wafercarrier}). Due to mechanical constraints between the probe card and the wafer clamping ring used to hold the wafer in place on the carrier during measurements, 264 out of  potentially functional 268  dice in each wafers can be probed with the cryoprobe\footnote{The 4 SiPMs per wafer not tested with the cryoprobe are not discarded, but instead used to build tiles subsequently tested to ensure compliance with the DS-20k tile specification, presented in a separate work.}.

The wafer carrier is placed in a semi-automatic loader. The loader is evacuated and transports the wafer (in vacuum) into the main chamber via a robot arm. The arm loads the wafer carrier on a gold-plated copper chuck that acts as a common cathode for current-voltage (IV) measurements. 

A liquid nitrogen (LN2) cooling coil inside the chuck brings its temperature down to 77~K. The LN2 that flows in the coil is regulated by a needle valve placed on an LN2 transport line that is used to connect the PAC200 probe station to an external LN2 storage Dewar. Two heater resistors maintain the chuck temperature constant to a precision of better than 0.5~K ($1 \sigma$) at 77~K.

The wafer carrier makes thermal and electrical contact with the chuck through four indium pads. The probe card needles touch down on the wafer surface, contacting the SiPM aluminum pads (1 needle per SiPM). 
Two probe cards with 24 and 8 needles, respectively, are used to test the wafers and periodically interchanged as part of regular system maintenance. An S500 Parametric tester equipped with Keithley 2612B Source Meters Units (SMU, 1 SMU channel per needle) is used to bias the SiPMs and read their current~\cite{S500}. The system noise floor (including cables, SMUs, and probe card) is around $10~\text{pA}$.

The Keithley Automated Characterization Suite (ACS) software program is used in conjunction with the FormFactor PAC200 Velox software to control the cryoprobe automatically, move the chuck from dice to dice, and execute the measurement sequence presented in Sec.~\ref{sec:expdetail}. A MIDAS-based slow control system \cite{MIDAS} is used to monitor the entire procedure and 
protect the system against human errors ({\it e.g.} wrong needle contact heights). Depending on the probe card configuration ({\it i.e.} number of needles) and the S500 measurement settings, at least 14 touchdowns are needed to test the accessible wafer surface (264 out of 268 SiPMs). The rate to probe a wafer, including the wafer loading, cooling, measurement, and unloading time, is ~0.35 wafers/hour. The cryoprobe is operated 12 hours per day, 7 days per week, with an up-time of 85\%, which results in a maximum wafer throughput of 25 wafers per week.

\section{Experimental Details}
\label{sec:expdetail}

\begin{figure}[ht]
\centering
\centering\includegraphics[width=0.99\linewidth]{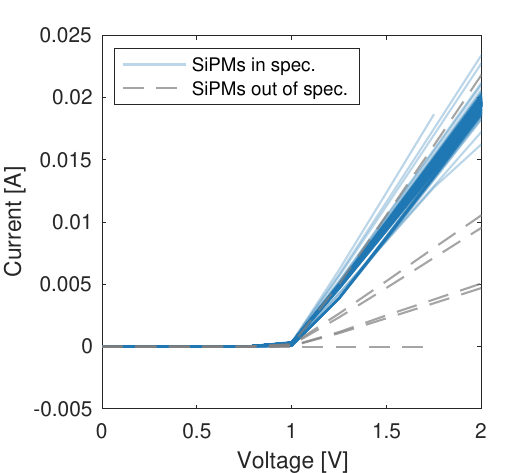}
\centering\includegraphics[width=0.99\linewidth]{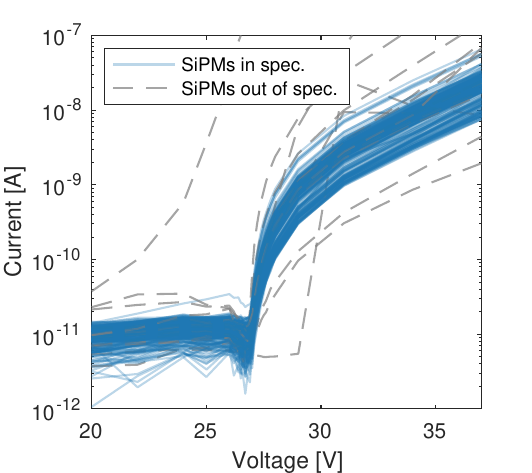}
\caption{Forward (top) and reverse (bottom) bias current-voltage (IV) curves for the production Wafer \WAFER of Lot \LOT. The dashed lines are IV curves that correspond to dice which were discarded based on the selection criteria presented in this work.}
\label{fig:IVs}
\end{figure}

The forward and reverse bias current-voltage (IV) curves of the testable SiPMs on a wafer (264 out of 268 SiPMs) are collected at 77~K. The cryoprobe vacuum chamber is not completely light-tight: there is a photon-induced rate on the order of a few $\text{kHz}$ per SiPM. No external light source is therefore needed to create a measurable current when collecting the reverse bias IV curves. Fig~\ref{fig:IVs} shows IV curves from one  production wafer.

\begin{figure}[ht]
\centering
\centering\includegraphics[width=0.99\linewidth]{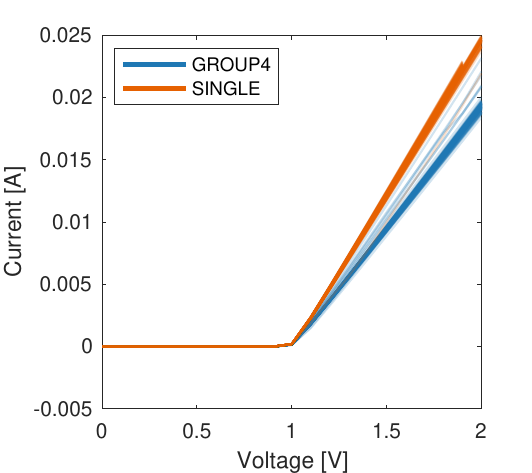}
\caption{Forward bias IV curves for one production wafer when simultaneously measuring 4 SiPMs in parallel (GROUP4) versus a single SiPM (SINGLE).}
\label{fig:IVs_slope}
\end{figure}

To reduce the testing time, SiPMs are measured in parallel. The maximum  number of SiPMs that can be measured simultaneously depends on two factors: (i) the probe card configuration ({\it i.e.} the number of needles on the card) and, (ii) the parasitic overall series resistance of the  measurement system, which includes the cabling and the cryogenic chuck responsible for inducing a voltage drop and therefore lowering the measured current. This drop is negligible for reverse bias IV curves, but it is relevant for forward bias IV curves due to the large sourced current (in the order of a few milliAmp), especially when grouping SiPMs, as shown in Fig.~\ref{fig:IVs_slope}\footnote{The total system resistance is estimated to be $3.7~\upOmega$, of which $2~\upOmega$ is due to the resistance between the chuck top surface and its hermetic bottom surface. This value was estimated by the change in slope of the forward bias IV curves when going from a single channel to a four-channel measurement, as shown in Fig.~\ref{fig:IVs_slope}. The extrapolated value was also confirmed with a dedicated measurement done on the cryogenic chuck of the PAC200 system.}.
We also observe a growing instability {\it i.e.} larger fluctuation, in the measured current even for reverse bias IV curves, as the number of SiPMs tested in parallel increases, which is probably due to the common cathode configuration.  Overall, we found that grouping 4 (8) SiPMs in parallel for forward (reverse) IV curves represents a good compromise between these effects and the testing throughput.

IV curves in forward bias are measured at constant voltage increments of 0.25~V up to 2~V, which corresponds to a forward current of about 20~mA. IV curves in reverse bias are measured starting from 20~V, several volts below the expected breakdown voltage at 77~K (27.2~V), up to 37~V, which corresponds to roughly 10~V of over voltage, larger than the maximum planned SiPM operating over voltage in the DS-20k detector (7~V). The voltage step is adjusted to be smaller in the range [26.4--27.6]~V to obtain a more accurate extrapolation of the breakdown voltage. The step choice is a compromise between accuracy and overall wafer testing time.

\begin{figure*}[ht]
\centering\includegraphics[width=0.8\linewidth]{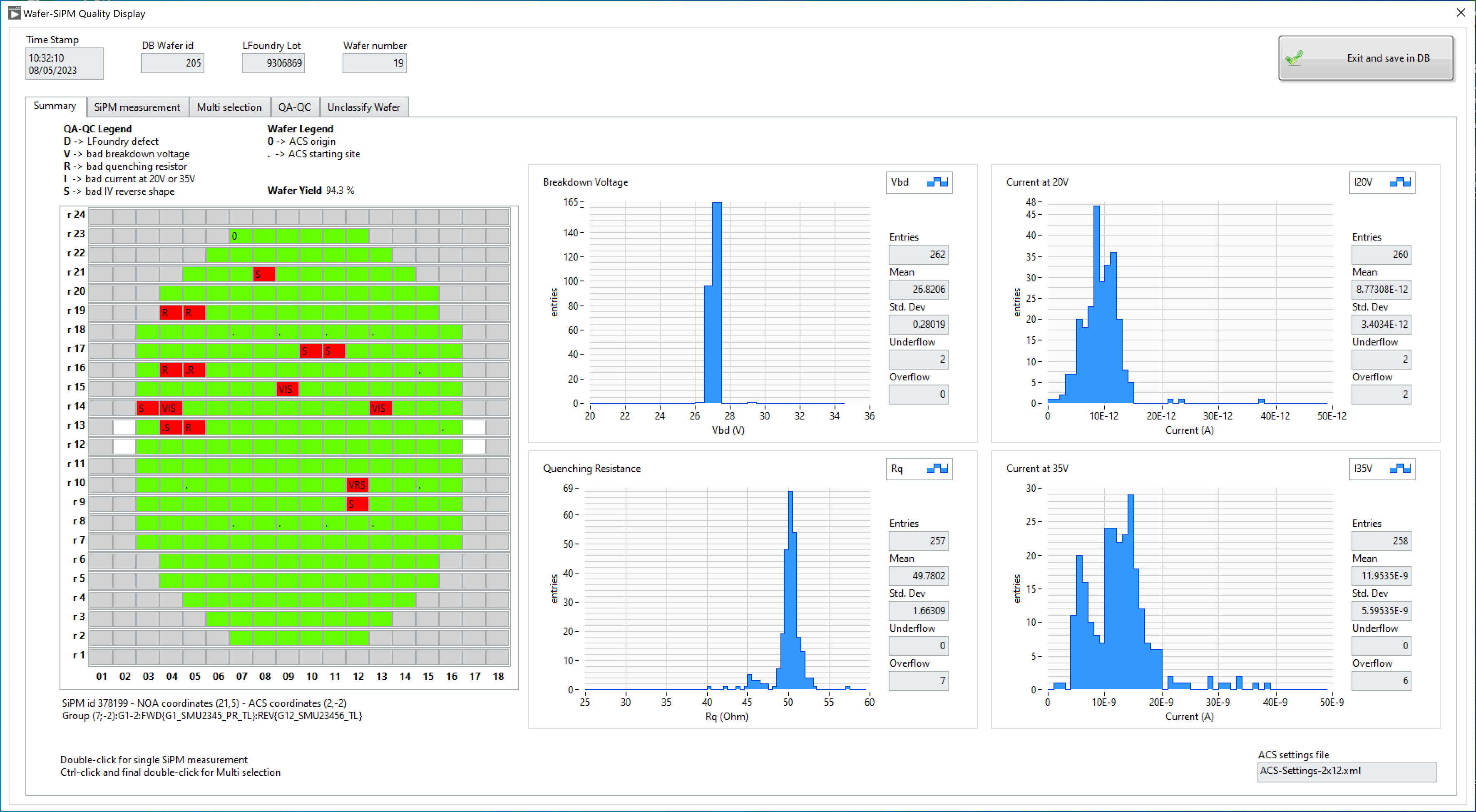}
\caption{Main display of the NOA Test Bench SiPM (NOA-TB-SiPM) Labview Application used to analyze the SiPM IV curves measured with the Keithley Automated Characterization Suite (ACS). For each wafer die a reverse and forward bias IV curve is measured at 77~K. Several results are shown to the operator before being stored in the database:  (top-left) Breakdown Voltage ($V_\mathrm{bd}$), (top-right) leakage current below the breakdown voltage (at 20~V) ($I_L$), (bottom-left) quenching resistor ($R_\mathrm{q}$), and (bottom-right) current at 35~V. Wafer dice coloured in green (red) pass (fail) all (at least one) the wafer-level requirements presented in this work.}
\label{fig:labview}
\end{figure*}

A Labview-based application, shown in Fig.~\ref{fig:labview}, was developed to analyze, display, and store the data from the Keithley ACS software (Sec.~\ref{sec:hardware}).
A PostgreSQL database hosted at INFN-CNAF (Italy) stores the analysis results, tracks the wafers during the entire process flow in NOA, and provides a comprehensive set of criteria for the QA/QC assessment and inventory purposes. The database is also used to keep track of the location and shipping status of wafers, which will be especially important as the detector is assembled.

\section{SiPM Acceptance Criteria}
\label{sec:qaqc}

The reverse and forward bias IV curves of each SiPM from all the production wafers are analyzed to ensure compliance with the following DS-20k wafer-level requirements (at 77~K): (i) breakdown voltage $V_\mathrm{bd}=27.2\pm1.0$~V, (ii) quenching resistor $R_\mathrm{q}=3.35\pm1.50~\text{M}\upOmega$, (iii) leakage current below breakdown (at 20~V) $I_L\le 40~\text{pA}$ and (iv) Goodness of Fit (GOF)~$\le 20$. The first three requirements are set to obtain an acceptable production yield based on the dispersion of the just introduced parameters measured on  pre-production FBK wafers. The goodness of fit parameter is instead based on a $\chi^2$ analysis of the shape of each reverse bias SiPM IV curve and places an upper limit on the SiPM correlated avalanche noise, {\it i.e.} crosstalk and afterpulse probability (Sec.~\ref{sec:GOF}). $V_\mathrm{bd}$, GOF, and $I_L$ are extracted from the reverse bias IV curve, while $R_\mathrm{q}$ is measured from the forward bias IV curve.

\subsection{Breakdown Voltage Distribution}
\label{sec:Vbd}

\begin{figure}[ht]
\centering
\includegraphics[width=0.99\linewidth]{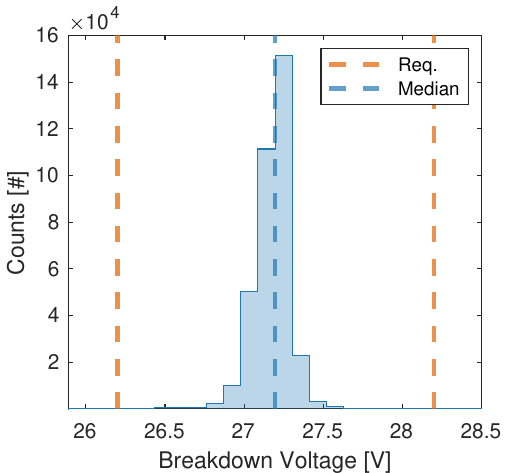}
\includegraphics[width=0.99\linewidth]{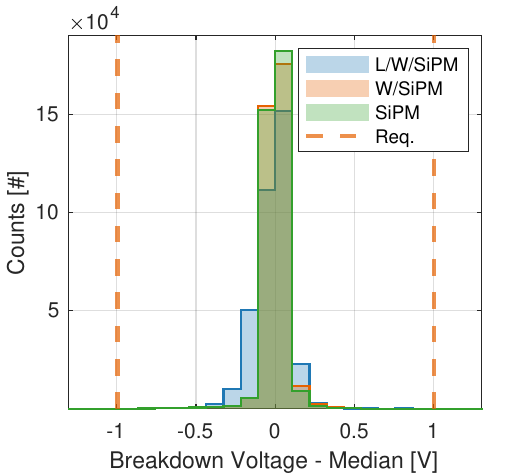}
\caption{(Top) Breakdown Voltage distribution of all the \NSIPM tested SiPMs. The red dashed lines represent the DS-20k requirements ($27.2\pm1.0$~V), while the blue line the histogram median. \NSIPMVBDout SiPMs are outside the range of the plot. (Bottom) Breakdown voltage variability within single wafers (SiPM), mixing wafers in the same Lot (W/SiPM) and mixing Lots (L/W/SiPM). W and L stand for Wafer and Lot. See text for more details.
}
\label{fig:hbVbd}
\end{figure}

\begin{figure*}[ht]
\centering\includegraphics[width=0.9\linewidth]{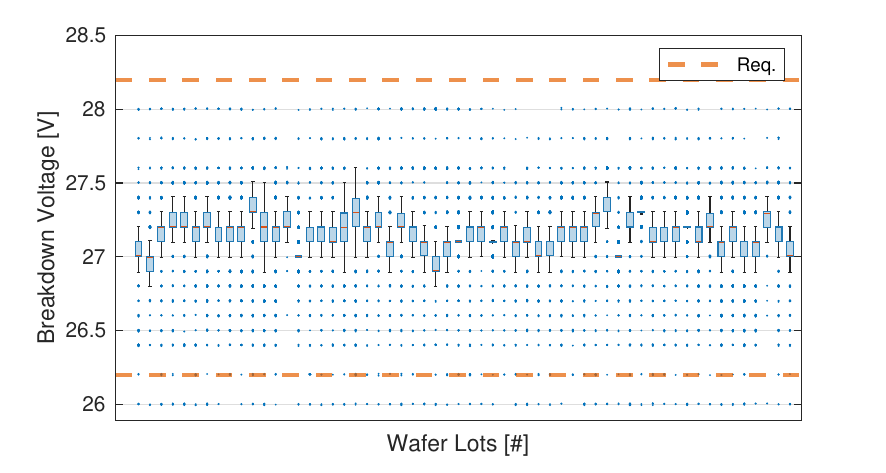}
\caption{Box plot of the measured breakdown voltage for the tested Lots (56 production Lots, 2 engineering Lots). The orange line inside each box represents the sample median, while the top and bottom edges are the upper and lower quartiles. Dots represent outliers more than $1.5$ InterQuartile Range (IQR) away from the top or bottom of each box. The dashed lines represent the DS-20k requirements ($27.2\pm1.0$~V).}
\label{fig:bVbd}
\end{figure*}

The SiPM breakdown voltage is defined as the voltage at which the first derivative  with respect to the voltage of the SiPM current (in log space) is at a maximum \cite{Nagai2018}. A narrow distribution of this parameter is crucial when grouping SiPMs in tiles (24 SiPMs, Sec.~\ref{sec:introduction}) since a large difference in breakdown voltages  would result in a gain and amplitude mismatch when operating tiles at a common bias, reducing the tile single PE resolution~\cite{Gallina2019}\footnote{Within DS-20k, the single photoelectron resolution is defined as the ratio between half of the Gaussian width of the first Photo-electron Equivalent (PE) peak and its centroid.}.\\Fig.~\ref{fig:hbVbd} (top) shows the breakdown voltage distribution for all the \NSIPM tested SiPMs, while Fig.~\ref{fig:bVbd} shows the same information grouping SiPMs for their production Lot. At $77$~K, the average breakdown voltage is measured to be $\overline{V}_\mathrm{bd}=27.19\pm0.05$~V\footnote{In what follows, we will always report summary results using the histogram median and half of its InterQuartile range (IQR/2) so that outliers don't affect the estimators.}. The dashed lines in the figure represent the DS-20k requirement ($27.2\pm1.0$~V). The tested SiPMs comfortably satisfy this requirement.

The variance of the breakdown voltage distribution $\sigma^2_{V_\mathrm{bd}}$ after exclusion of the out-of-spec data is 0.019~$\text{V}^2$. This quantity can be decomposed into components:
\begin{equation}
\sigma^2_{V_\mathrm{bd}}=\sigma^2_{V_\mathrm{bd}/\text{Lot}}+\sigma^2_{V_\mathrm{bd}/\text{Wafer}}+\sigma^2_{V_\mathrm{bd}/\text{SiPM},}
\end{equation}
where $\sigma^2_{V_\mathrm{bd}/\text{Lot}}$ is the Lot-to-Lot variance, $\sigma^2_{V_\mathrm{bd}/\text{Wafer}}$ is the Wafer-to-Wafer variance of a single Lot, and $\sigma^2_{V_\mathrm{bd}/\text{SiPM}}$ is the SiPM-to-SiPM variance of a single wafer. Each variance contribution has been estimated with a Variance Component Analysis (VCA) \cite{Searle-1992}. The results are summarized in Table~\ref{T:VCA_Vbd}. 55.9\% of the variance is due to the SiPM-SiPM variability within a wafer. The Lot contribution to the variance is  36.9\%, and the remaining 7.3\% is due to the Wafer to Wafer variability.

\begin{table}[ht]
\centering
\begin{tabular}{cccccc}
   \toprule
Factor & Samples & VC & \% of Tot. & Cum. VC & Cum. Fact.\\
\midrule
 SiPM & \NSIPM & 10.8 & 55.9 & 10.8 & SiPM\\ 
 Wafer & 1360 & 1.41 & 7.3 & 12.21 & W/SiPM\\ 
 Lot & 58 & 7.13 & 36.9 & 19.34 & L/W/SiPM\\ 
\midrule
 Total & & 19.34 & 100 & & \\ 

    \bottomrule
\end{tabular} 
\caption{Variance Component Analysis (VCA) of the Breakdown Voltage ($V_\mathrm{bd}$) distribution. VC is the Variance Component in $\text{V}^2$ x 1000; \% of Tot. is the percentage of each variance component; Cum. VC is the cumulative variance considering the incremental contribution of each component; Cum. Fact. represents the incremental nested factors contributing to the cumulative variance (W and L stand for Wafer and Lot).
}
\label{T:VCA_Vbd}
\end{table}

\begin{figure}[ht]
\centering
\centering\includegraphics[width=0.9\linewidth]{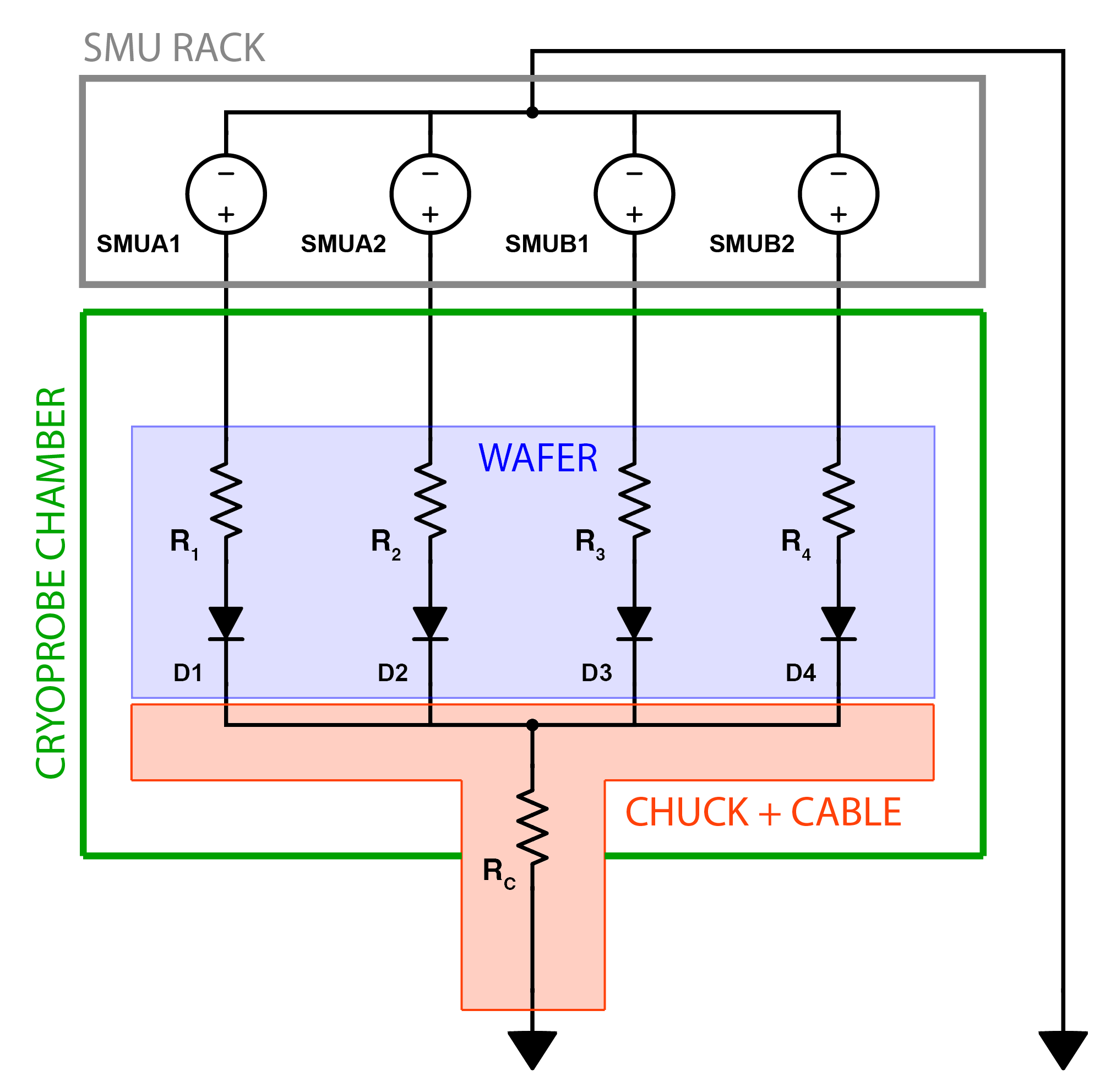}
\caption{Schematic of the measurement system for forward bias IV curves. $R_c$ is the series resistance of the cryoprobe+SMUs measurement system, measured to be $3.7~\upOmega$. $R_i$ with $i={1,2,3,4}$ are the resistance of the 4 SiPMs measured in parallel.}
\label{fig:voltagedrop}
\end{figure}

Fig.~\ref{fig:hbVbd} (bottom) shows a visual representation of this analysis using histograms. 

We present three histograms: L/W/SiPM, W/SiPM, and SiPM. The L/W/SiPM histogram was obtained by subtracting the median value of the entire distribution from the breakdown voltage of each individual SiPM. Essentially, this represents the whole distribution, shifted to align its median with zero, including all three contributions (Lot, Wafer, SiPM) to the variance. The W/SiPM histogram was created by subtracting the median breakdown voltages within each Lot from the individual SiPM breakdown voltages. This approach removes the Lot-to-Lot contribution to the variance, resulting in each Lot having a median of \SI{0}{\volt}. Similarly, the SiPM histogram was produced by subtracting the median breakdown voltages within each Wafer, thus eliminating the Wafer-to-Wafer contribution to the variance. Here the only contribution to the variance is due to the SiPMs themselves.

This representation shows that the breakdown voltage variability drops drastically when the contribution of the Lot is removed. Specifically, moving from L/W/SiPM to W/SiPM, the variance decreases by 36.9\% of the total variance, according to Table~\ref{T:VCA_Vbd}. Furthermore, when the Wafer-to-Wafer variability is removed, {\it i.e.}, comparing W/SiPM with SiPM, the variance reduction is 1.41 $\times 10^{-3}$ \( \text{V}^{2}\) (7.3\% of the total), smaller than the previous case, resulting in a residual variance of 10.8 $\times 10^{-3}$ \( \text{V}^{2}\) that accounts for the 55.9\% of the total.

\begin{figure}[ht]
\centering
\includegraphics[width=0.99\linewidth]{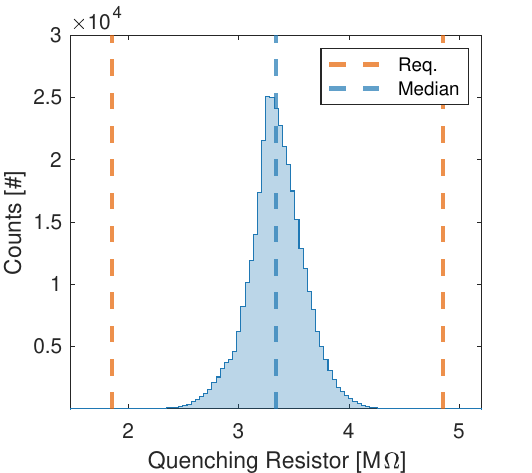}
\includegraphics[width=0.99\linewidth]{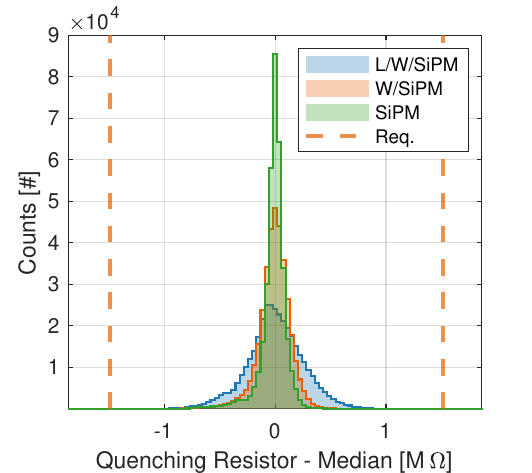}
\caption{(Top) Quenching Resistor distribution of all the \NSIPM tested SiPMs. The red dashed lines represent the DS-20k requirements ($3.35\pm1.50~\text{M}\upOmega$), while the blue line the histogram median. \NSIPMRQout SiPMs are outside the range of this plot. (Bottom) Quenching Resistor variability within single wafers (SiPM), mixing wafers in the same Lot (W/SiPM) and mixing Lots (L/W/SiPM). W and L stand for Wafer and Lot. See text for more details.
}
\label{fig:hbRq}
\end{figure}

\subsection{Quenching Resistor Distribution}
\label{sec:Rq}

The SiPM quenching resistor is calculated from the linear part of each forward biased IV curve (Fig.~\ref{fig:IVs}) in the regime where the resistance of the SPAD's pn-junction is negligible ($>1.25$~V)~\cite{Otte2016}. The inverse of the IV curve slope, after correcting for the voltage drop of the measurement system (Sec.~\ref{sec:hardware}), is the parallel resistance of all the SiPM SPADs' quenching resistors. A narrow distribution of this parameter is desirable when grouping SiPMs in tiles since a mismatch in the SiPM quenching resistor would produce a pulse-to-pulse variability of the SiPM pulse shape, depending on the triggered SiPM. If $R_c$ is the series resistance of the measurement system ($R_c=3.7~\upOmega$, Sec.~\ref{sec:expdetail}), the SiPM SPAD quenching resistor ($R_\mathrm{q}$) can be derived as follows
\begin{equation}
\label{eq:quenching}
R_\mathrm{q}=[R^*-4R_c]\times N_{\text{cell}}
\end{equation}
where $R^*$ is the resistance measured using a linear fit to the forward bias IV-curve and $N_{\text{cell}}=94904$ is the number of SPAD cells in each SiPM. The factor 4 in Eq.~\ref{eq:quenching} accounts for the fact that SiPMs are measured in parallel groups of 4, as introduced in Sec.~\ref{sec:hardware} and also shown in Fig.~\ref{fig:voltagedrop}.

\begin{figure*}[ht]
\centering\includegraphics[width=0.9\linewidth]{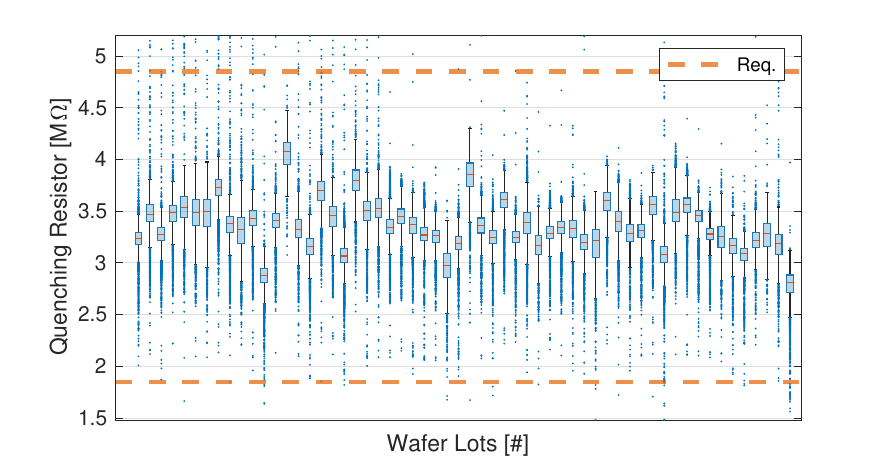}
\caption{Box plot of the measured quenching resistor for the tested Lots (56 production Lots, 2 engineering Lots). The line inside each box represents the sample median while the top and bottom edges are the upper and lower quartiles. Dots represent outliers more than $1.5$ InterQuartile Range (IQR) away from the top or bottom of each box. The dashed lines represent the DS-20k requirement ($3.35\pm1.50~\text{M}\upOmega$).}
\label{fig:bRQ}
\end{figure*}

Fig.~\ref{fig:hbRq} (top) shows the quenching resistor distribution for all \NSIPM SiPMs tested, while Fig.~\ref{fig:bRQ} shows the same information grouping SiPMs for their production Lot. At $77$~K, the average quenching resistor is measured to be $\overline{R}_\mathrm{q}=3.34\pm 0.15$~$\text{M}\upOmega$. The dashed lines represent the DS-20k requirements ($3.35\pm1.50~\text{M}\upOmega$). Table~\ref{T:VCA_Rq} reports the quenching resistor VCA analysis. The production variance ($\sigma^2_{R_\mathrm{q}}$) after removing out-of-spec devices is  equal to 0.073~(M$\upOmega^2$). 
\begin{table}[ht]
\centering
\begin{tabular}{cccccc}
   \toprule
Factor & Samples & VC & \% of Tot. & Cum. VC & Cum. Fact.\\
\midrule
 SiPM & \NSIPM & 0.019 & 26.3 & 0.019 & SiPM\\
 Wafer & 1360 & 0.009 & 11.9 & 0.028 & W/SiPM\\
 Lot & 58 & 0.045 & 61.8 & 0.073 & L/W/SiPM\\
\midrule
 Total &  & 0.073 & 100 & &\\ 
    \bottomrule
\end{tabular} 
\caption{Variance Component Analysis (VCA) of the Quenching Resistor ($R_\mathrm{q}$) distribution. VC is the Variance Component in $(\text{M}\upOmega)^2$; \% of Tot. is the percentage of each variance component; Cum. VC is the cumulative variance considering the incremental contribution of each component; Cum. Fact. represents the incremental nested factors contributing to the cumulative variance (W and L stand for Wafer and Lot).
}
\label{T:VCA_Rq}
\end{table}

Fig.~\ref{fig:hbRq} (bottom) shows a visual representation of this analysis, similarly to what was done in Sec.~\ref{sec:Vbd} for the SiPM breakdown voltage. The variability of the quenching resistor drops drastically when the contribution of the Lot is removed. Specifically, moving from L/W/SiPM to W/SiPM, the variance decreases by \(0.045 \, (\text{M}\upOmega)^{2}\), or 61.8\% of the total variance, according to Table~\ref{T:VCA_Rq}. Additionally, when the Wafer-to-Wafer variability is removed, {\it i.e.}, comparing W/SiPM with SiPM, the variance is reduced by \(0.009 \, (\text{M}\upOmega)^{2}\) which is a smaller amount compared to the previous case and corresponding to 11.9\% of the total variance. The residual variance (SiPM) is 26.3\% of the total.

\subsection{Spatial Distribution of the Breakdown Voltage and Quenching Resistor}

\begin{figure}[ht]
\centering
\includegraphics[width=0.99\linewidth]{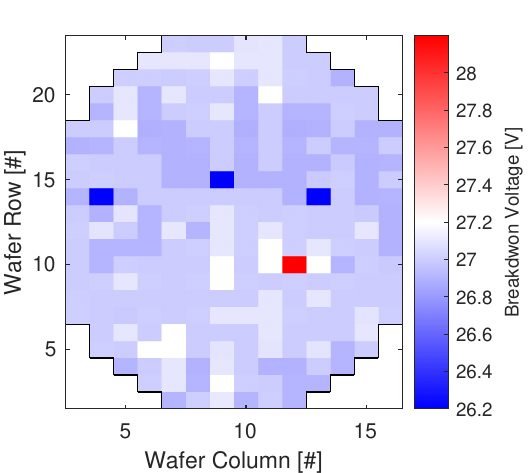}
\includegraphics[width=0.99\linewidth]{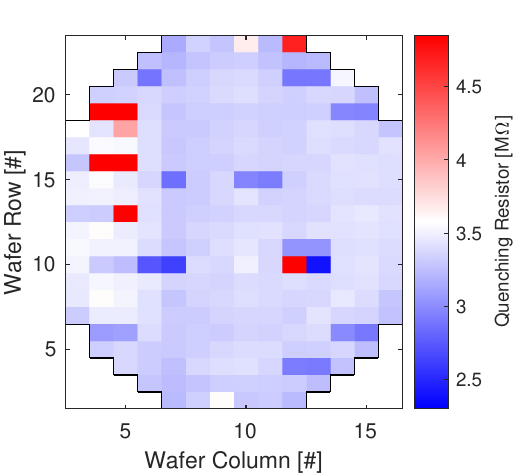}
\caption{Spatial distribution of the Breakdown Voltage (top) and Quenching Resistor (bottom) for Wafer~\WAFER of Lot~\LOT. This wafer is the same whose IV curves are shown in Fig.~\ref{fig:IVs}. To keep the scale readable, we associated to the few dice whose breakdown voltage and quenching resistor is outside the range of the figure a value equal to the closest axis range boundary.}
\label{fig:heatmpas}
\end{figure}

In the previous sections, we presented the variability in breakdown voltage and quenching resistance using histograms and VCA analyses. Fig.~\ref{fig:heatmpas} shows the typical spatial distribution of these two parameters on a representative production wafer. The results indicate that both parameters are generally uniform across the wafer, with only a few dice falling outside of specifications. Note that this wafer was selected solely for illustrative purposes; the extremely narrow distributions of breakdown voltage and quenching resistance within individual wafers were quantitatively confirmed in the previous sections. 

Based on this statistical analysis, the current strategy at the tile assembly stage is to freely combine dice from the same wafer and, when necessary, include dice from wafers within the same Lot. Given the consistent performance of SiPMs across a Lot, it is reasonable to treat the Lot as the fundamental grouping element for tile assembly rather than the individual wafer.

\subsection{Correlated and uncorrelated noise}

Dark and correlated avalanche noise are critical SiPM parameters that affect the overall DS-20k detector performance by increasing the number and the fluctuations of detected photons. Fluctuations in the number of detected photo-electrons, for example, may decrease the energy resolution of the detector, as shown in Ref.~\cite{Gallina2022}.

Dark noise is a spontaneous charge signal generated by an electron-hole pair formed by thermal or field-enhanced processes \cite{Hurkx1992}. Correlated Avalanche (CA) noise is due to at least two processes: the production of secondary photons in the gain amplification stage during primary avalanches and the trapping and subsequent release of charge carriers produced in avalanches (afterpulsing). Several techniques are used to characterize these two sources of noise while operating the SiPM in pulse counting mode~\cite{Ako,Gallina2019,gallina2021development}. However, estimating these properties directly from the wafer-level reverse bias IV curves is challenging.

In Sec.~\ref{sec:GOF}, we introduce a new technique that, starting from the shape of the reverse bias IV curve and, regardless of the level of SiPM illumination, ensures that all the production-graded SiPMs: (i) have similar CA noise and (ii) don't present an out-of-spec voltage dependence ({\it e.g.} high DCR). The leakage current before the breakdown voltage is also used as a control parameter to identify out-of-spec SiPMs ({\it e.g.} shorts). More details are reported in Sec.~\ref{sec:Il}.

\subsubsection{Goodness of fit}
\label{sec:GOF}
To reach the DS-20k design performance, all production-graded SiPMs must have similar correlated noise. The reverse bias SiPM IV curve is sensitive to this quantity. In general, the SiPM current under illumination can be written as~\cite{Gallina2022}
\begin{linenomath}
\begin{equation}
\label{eq:PDE_INFN}
I(V,\lambda)=f(V)\times\Big[\text{PDE}_{\lambda}(V)\times\Phi(\lambda)+\text{R}^\text{DCR}_{\text{SiPM}}(V)\Big]
\end{equation}
\end{linenomath}
where $V$ is the SiPM reverse bias voltage, $\text{PDE}_{\lambda}$ is the SiPM Photon Detection Efficiency (PDE), and $\Phi(\lambda)$ is the photon flux. $\text{R}^\text{DCR}_{\text{SiPM}}(V)$ is the SiPM DCR. $f(V)$ is a correction factor that can be written as 
\begin{equation}
\label{eq:fv}
    f(V)\sim q_e\times\big(1+\overline{\Lambda}\big)\times \overline{\text{G}}_{1\text{ PE}},
\end{equation}
where ($\overline{\text{G}}_{1\text{ PE}}$) is the SiPM gain for 1 Photo-electron Equivalent (PE) charge, ($\overline{\Lambda}$) is the CA noise contribution that artificially increases the total current produced by the SiPM and ($q_e$) is the electron charge~\cite{Gallina2022}\footnote{The SiPM gain ($\overline{\text{G}}_{1\text{ PE}}$) and the CA noise contribution ($\overline{\Lambda}$) are voltage dependence as shown, for example, in Ref.~\cite{Gallina2019,Gallina2022}}.

$f(V)$ is a function of the applied bias voltage $V$ but can be considered wavelength-independent because it depends only on the SiPM intrinsic characteristics. This is because the average sensor gain ($\overline{\text{G}}_{1\text{ PE}}$) depends on the SiPM SPAD capacitance while afterpulses and optical crosstalk that contribute to the sensor CAs depend on impurities and cell geometry.

At 77~K, the DCR of these SiPMs is on the order of $\text{mHz}/\text{mm}^2$ and makes a negligible contribution to the total current~\cite{Gola2019}. If we assume that all the SiPMs have identical characteristics {\it e.g.} efficiency and correlated noise, and their DCR  is negligible 
compared to the incident photon flux $\Phi$, the ratio point-to-point ({\it i.e.} at the same bias voltage) of two SiPMs' IV curves as a function of the applied bias voltage $V$ is a voltage-independent quantity, only proportional to the ratio of the photon fluxes seen by the SiPMs, here called $k$, as follows 
\begin{equation}
\label{eq:ratio_IVs}
\frac{I_1(V,\lambda)}{I_2(V,\lambda)}=\frac{\Phi_1(\lambda)}{\Phi_2(\lambda)}\equiv k.
\end{equation}

\begin{figure}[ht]
\centering
\includegraphics[width=0.99\linewidth]{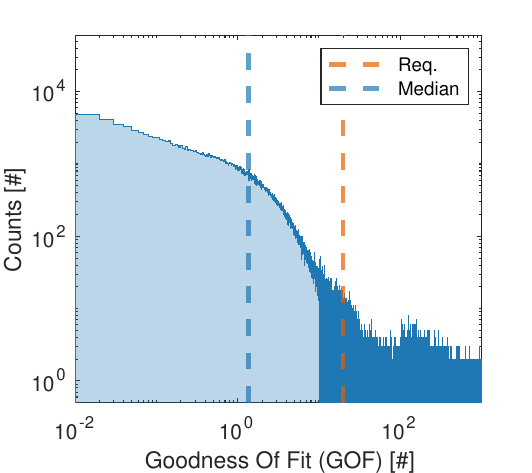}
\caption{Goodness of Fit (GOF) distribution of all the \NSIPM SiPMs tested. The dashed line represents the DS-20k requirements ($\text{GOF}\le 20$), while the blue line the histogram median. \NSIPMGOFout SiPMs are outside the range of the plot.}
\label{fig:hChi2}
\end{figure}

In order to ensure compliance of the wafer level IV curves with Eq.~\ref{eq:ratio_IVs}, we introduce a parameter called goodness of fit defined as follows 
\begin{equation}
\text{GOF}=\mathlarger{\mathlarger{\sum_i}}\frac{(I_i-k\overline{I_i})^2}{\overline{\sigma^2_i}(N-1)},
\label{eq:GOF}
\end{equation}
where $N$ is the number of measured points, $I_i$ is the IV curve of any SiPMs in the wafer, and $\overline{I_i}$ is current of a SiPM chosen as to be the reference for the entire DS-20k production, and whose IV curve is named reference-IV. $\overline{\sigma_i}$ is instead the uncertainty of the reference IV curve. It is set to be equal to 22\% and it was computed accounting for the average point-to-point fluctuation. The sum in Eq.~\ref{eq:GOF} is done over the bias voltage points in the IV curve, starting  0.7~V away from the breakdown voltage of the reference IV-curve ($\overline{I}$) after alignment of the breakdown voltage of the curve $I$ with the one of $\overline{I}$. This was done to avoid the steep IV curve rise, close to the breakdown voltage, where large fluctuations due to baseline noise are present and would affect the GOF computation.\\The scaling factor $k$ (Eq.~\ref{eq:ratio_IVs}) can be calculated analytically by minimizing Eq.~\ref{eq:GOF} as follows
\begin{equation}
\label{eq:k}
k=\frac{\sum_i(\overline{I_i}~ I_i)/\overline{\sigma^2_i}}{\sum_i{\overline{I^2_i}/\overline{\sigma^2_i}}}.
\end{equation}
Fig.~\ref{fig:hChi2} shows a  histogram of the GOF parameter for all the \NSIPM tested SiPMs. 

\begin{figure}[ht]
\centering
\centering\includegraphics[width=0.99\linewidth]{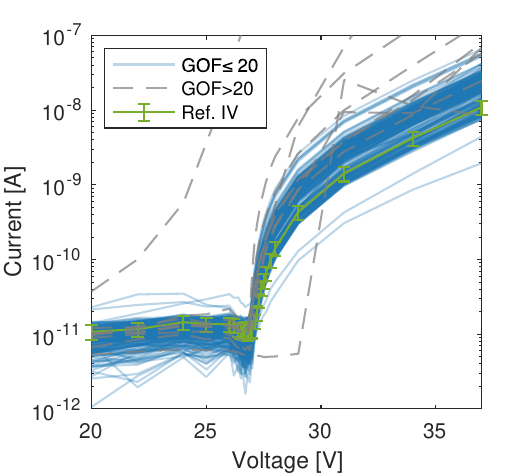}
\centering\includegraphics[width=0.99\linewidth]{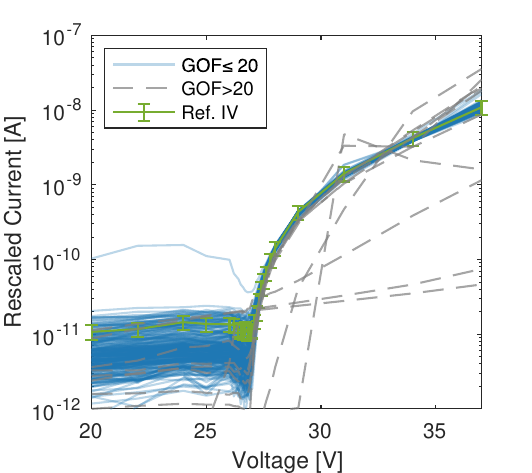}
\caption{
Reverse bias IV curves for the production Wafer \WAFER of Lot \LOT categorized according to their Goodness of Fit (GOF) parameter (Top). The bottom figures shows the same IVs, always categorized according to their GOF and scaled for the corresponding $k$-factor, computed as per Eq.~\ref{eq:k}.}
\label{fig:IVChi2}
\end{figure}

The average GOF is $1.36\pm1.49$ meaning all SiPMs IV curves are in good agreement with the reference IV curve when scaled with a constant scaling factor. For instance the top panel of Fig.~\ref{fig:IVChi2} displays the IV curves from the wafers shown in Fig.~\ref{fig:IVs}, categorized by their GOF parameter. In contrast, the bottom panel of Fig.~\ref{fig:IVChi2} presents the same IV curves after scaling using the $k$-factor calculated from Eq.~\ref{eq:k}. This scaling compensates for fluctuations due to varying light fluxes beyond the breakdown voltage, resulting in most of the SiPM IV curves matching the reference one within its error margins. 

While IV curves with a $\text{GOF}>20$ deviate noticeably from the reference, there is minimal practical difference among curves with a $\text{GOF}\le20$, which is the requirement set for DS-20k production (as illustrated in Fig.~\ref{fig:hChi2}). The upper limit for the GOF parameter ($\text{GOF}\le 20$) was established at the 3$\sigma$ level, meaning that 99.87\% of the data satisfy the specification in a one-tailed test; indeed, a theoretical $\chi^2$ distribution with 5 degrees of freedom contains 99.87\% of its data below 20.

\subsubsection{SiPM leakage current}
\label{sec:Il}

The cryoprobe vacuum chamber is not light-tight (Sec.~\ref{sec:expdetail}), which means the SiPM current above the breakdown voltage can not be used as a measurement of the SiPM DCR. However, the current below breakdown (not multiplied contribution) can be used to identify out-of-spec SiPMs.\\Below the breakdown voltage, at 77~K, the SiPM current is proportional to the SiPM photon flux. In the absence of gain (Geiger mode multiplication M=1) or under the presence of a proportional gain ($ \text{M}<<\overline{\text{G}}_{1\text{ PE}}$) the small SPAD capacitance results in a very small current, on the order of a few $\text{fA}$ or less, even under kHz of external illumination. This is below the sensitivity of the measurement system.

\begin{figure}[ht]
\centering
\includegraphics[width=0.99\linewidth]{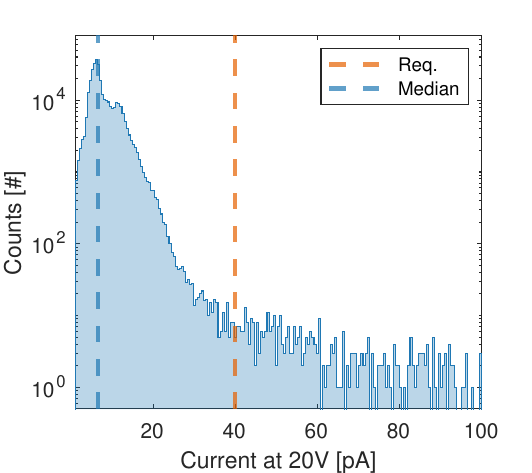}
\caption{Leakage Current distribution at 20~V  of all the \NSIPM tested SiPMs. The dashed line represents the DS-20k requirements ($I_L\le 40~\text{pA}$), while the blue line the histogram median. \NSIPMIout SiPMs are outside the range of the plot.}
\label{fig:hI20}
\end{figure}

Fig.~\ref{fig:hI20} shows a histogram of the leakage current measured at 20~V, several volts below the average breakdown voltage (Sec.~\ref{sec:Vbd}).

At 77~K, the average leakage current is measured to be $\overline{I}_L=6.6\pm2.2~\text{pA}$,  which represents the measurement system noise floor (Sec.~\ref{sec:hardware}).  The vertical line represents the DS-20k requirement ($I_L\le 40~\text{pA}$). The upper spec limit for this parameter value was set at the 4$\sigma$ level: {\it i.e.} 99.99\% of the data meet the specs.   

\section{Correlation of the 300~K and 77~K Breakdown Voltage and Quenching Resistor data}

In Sec.~\ref{sec:expdetail}, we described the NOA wafer testing workflow, which achieves a throughput of approximately 0.35 wafers per hour. The entire wafer testing campaign for DS-20k—now nearly complete—took roughly a year and a half.

Next-generation noble liquid cryogenic experiments, such as Argo and Darwin, plan to have SiPM-based photon-sensitive areas in excess of $100~\text{m}^2$~\cite{Agnes2021,Aalbers2016}. Cryogenic wafer-level SiPM QA/QC at this scale will be challenging and will likely require several cryoprobing systems. 

In this section, we study the correlation of the dice failure rate at 300~K and 77~K, to understand if room-temperature testing is adequate. Room temperature measurements could likely be performed by the wafer supplier or using measurement systems that are more affordable than the cryogenic prober used for this work. 

\begin{figure}[ht]
\centering
\includegraphics[width=0.99\linewidth]{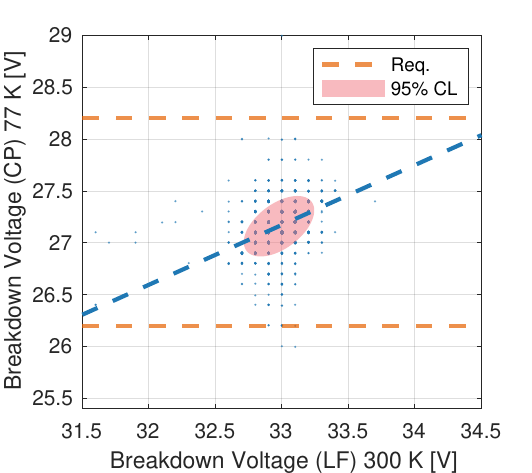}
\includegraphics[width=0.99\linewidth]{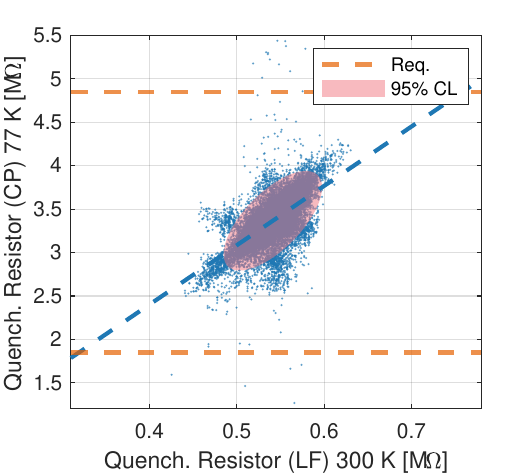}
\caption{Correlation of the breakdown voltage (top) and the quenching resistor (bottom) data measured by LFoundry (LF) at 300~K and by the cryoprobe (CP) at 77~K. Each point represents a SiPM. Horizontal orange lines are the SiPMs requirement at 77~K. Horizontal orange lines are the SiPMs requirement at 77~K, while the dashed blue lines are fit to the data. The red circle represents the 95~\% Confidence Level (CL) interval.}
\label{fig:correlation_Vbd_Rq}
\end{figure}

Fig.~\ref{fig:correlation_Vbd_Rq} shows the correlation between the SiPMs breakdown voltage (top) and the quenching resistor (bottom) measured by LFoundry at 300~K and by the cryoprobe system at 77~K. The cryoprobe (Sec.~\ref{sec:hardware}) in its actual configuration cannot in fact perform measurements at 300~K due to the mechanical tolerances of the probe card\footnote{The probe card is designed to operate at 77~K. At room temperature, the probe card expands and its needles do not contact the SiPM pads.}. Each point in the figure represents a SiPM.

For each production wafer, LFoundry tested the breakdown voltage of 10\% of the SiPM dice using a technique similar to the one presented in Sec.~\ref{sec:Vbd}. The quenching resistor was instead inferred by the measurement of the sheet resistance on nine test structures distributed over the wafer, using the Van Der Pauw method, placed on the wafer dicing street\footnote{The LFoundry measurements were performed before the deposition of the gold back-metallization (Fig.~\ref{fig:wafercassette}). The large backside resistance at this step of the manufacturing process prevents the usage of the forward bias IV curve, as was shown in Sec.~\ref{sec:Rq}. The two techniques yield comparable results, as shown in Ref.~\cite{pauw1958method}.}. Overall 44198 and 111316 dice were tested by LFoundry for $V_\mathrm{bd}$ and $R_\mathrm{q}$, respectively. The horizontal orange lines in Fig.~\ref{fig:correlation_Vbd_Rq} represent the  77~K  dice requirements (Sec.~\ref{sec:expdetail}) and the dashed blue lines are fit to the data. In the same figures, we also reported the 95~\% Confidence Level (CL) interval.
Table~\ref{T:Spec_Vbd_Rq_spec} reports the intersection between the 77~K requirements and the data fit line that shows a correlation between the  300~K  and the 77~K dataset. 

\begin{table}[ht]
\centering
\begin{tabular}{ccc}
   \toprule
Quantity & 77 [K] &  300 [K] \\
 \midrule
 $V_\mathrm{bd}$[V] & $27.2\pm1.0$ &  $33.1\pm 1.7$ \\
$R_\mathrm{q}$[M$\upOmega$] & $3.35\pm1.50$ & $0.54\pm0.22$  \\
    \bottomrule
\end{tabular} 
\caption{Breakdown voltage and quenching resistor requirements at 77~K and the extrapolated requirements at 300~K.}
\label{T:Spec_Vbd_Rq_spec}
\end{table}

\begin{figure}[ht]
\centering
\includegraphics[width=0.99\linewidth]{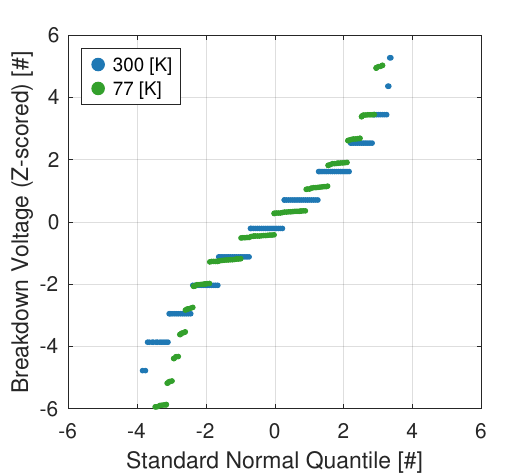}
\includegraphics[width=0.99\linewidth]{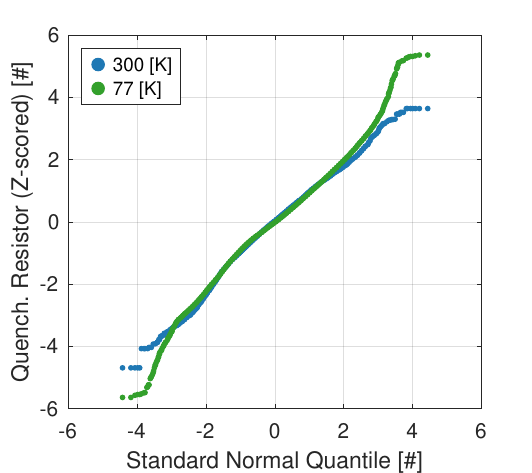}
\caption{QQ-Plots of the Z scored Breakdown Voltage (top) and Quenching Resistor (bottom) distributions for the 300~K and 77~K data. Both curves (300~K and 77~K) produce approximately a straight line  and almost perfectly overlap, apart from the edges, confirming that the two datasets follow the same probability distribution .}
\label{fig:qqplots}
\end{figure}

Other than being correlated, the two data sets have similar distributions. This can be seen by comparing the Z-Scored data of the two quantities at both temperatures using QQ-Plots~\cite{qqplot}, as shown in Fig.~\ref{fig:qqplots}. 
Given a variable $X={V_\mathrm{bd},R_\mathrm{q}}$, the corresponding Z-scored quantity is defined as
\begin{equation}
    \text{Z}=\frac{X-\overline{X}}{\sigma_X}
\end{equation}
where $\overline{X}$ and $\sigma_X$ are the mean and sigma of the corresponding distribution.\\Both curves (300~K and 77~K) produce approximately a straight line and almost perfectly overlap, apart from the very edges, confirming they follow the same probability distribution.\\The results presented in this section show that based on the measured data  more than 99\% of the dice that comply at room temperature are also compliant at cryogenic temperature. The low statistics of failing dice however do not allow us to conclude that the room temperature screening is sufficient to identify failing dice at cryogenic temperature. This is a consequence of the high wafer yield and low statistics of failing dice of the LFoundry production. New measurements, with a newly designed probe card that can operate at 300~K, are now being planned to further expand the results presented in this section. These also include the possibility of building a correlation not only based on the SiPMs breakdown voltage and quenching resistor, but also based on the GOF and leakage current parameters that could not be tested using the data presented in this work since LFoundry only performed measurements close to the dice breakdown voltage.

\section{Wafer Production Yield}
\label{sec:yield}

\begin{figure}[ht]
\centering
\includegraphics[width=0.99\linewidth]{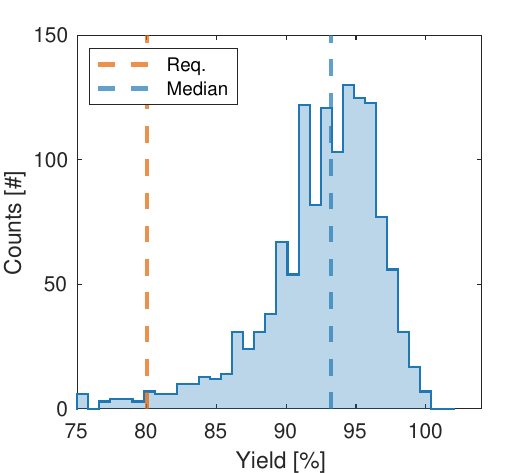}
\caption{Histogram of the wafer production yield after applying the screening presented in this work.  The two dashed lines
represent the histogram median (blue) and the expected production yield (red, 80\%), which is based on the yields of pre-production from FBK.}
\label{fig:hYield}
\end{figure}

\begin{figure*}[ht]
\centering\includegraphics[width=0.9\linewidth]{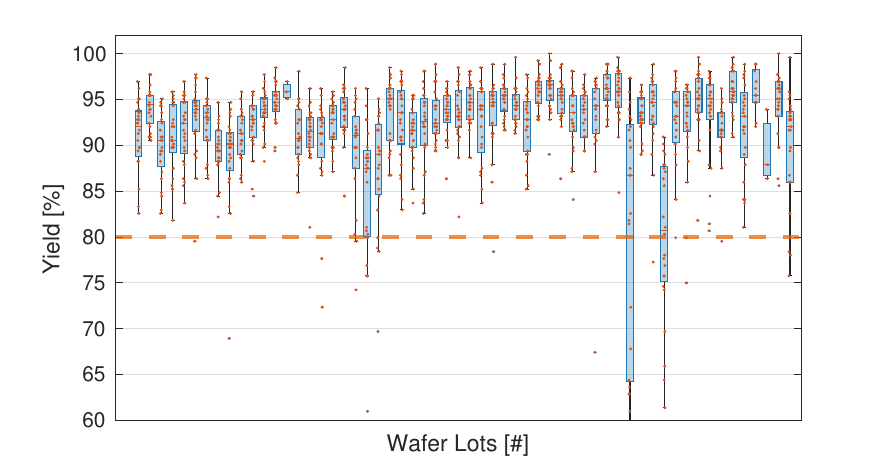}
\caption{Box plot of the yield for the tested Lots (56 production Lots, 2 engineering Lots). The line inside each box represents the sample median while the top and bottom edges are the upper and lower quartiles. Dots represents wafers in each Lot. The dashed lines represent the DS-20k requirement.}
\label{fig:byield}
\end{figure*}

Fig.~\ref{fig:hYield} shows a histogram of the wafer yield for all tested wafers.
The yield has been computed based on the 264 testable dice per wafer. The dashed orange line represents the assumed production yield (80\%), computed from pre-production wafers from FBK. 
\begin{figure}[ht]
\centering
\includegraphics[width=0.99\linewidth]{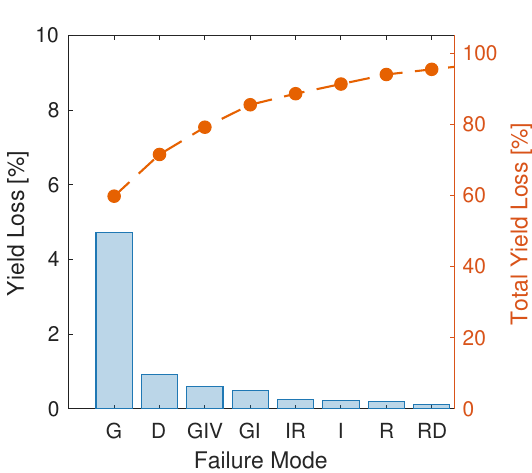}
\caption{Pareto plot that shows the source of the wafer yield loss differentiating SiPMs according to their failure loss mechanism: (i) Goodness of fit (G), (ii) leakage current below breakdown (I), (iii) defective pads (D), (iv) quenching resistor (R), (v) breakdown voltage (V)  and their combination. To keep the figure readable only the largest source of yield loss are considered, since they overall constitute 95\% of screened SiPMs.
}
\label{fig:hpareto_2}
\end{figure}
The tested wafers not only meet but significantly exceed this requirement, with an overall yield of $93.2\pm2.5$\%. The Lot-to-Lot yield variation is also mostly negligible, apart for few Lots, as shown in Fig.~\ref{fig:byield}. 

This represents a significant improvement if compared with the pre-production results. Indeed, high-volume CMOS foundries typically operate numerous parallelized pieces of equipment in a fully automated manner, which helps reduce non-uniformities and boost yield.

\begin{figure}[ht]
\centering\includegraphics[width=0.9\linewidth]{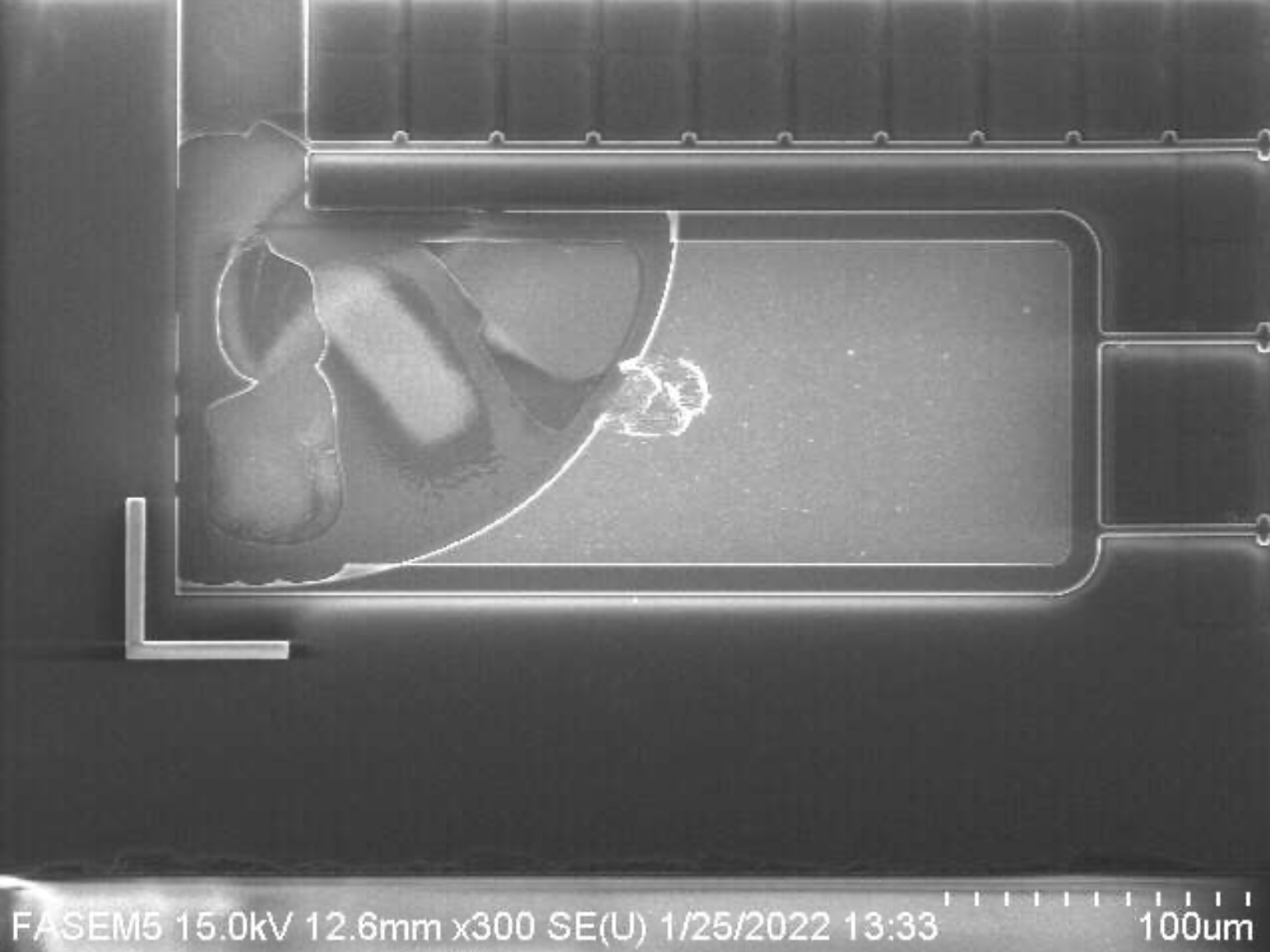}
\caption{Scanning Electron Microscope (SEM) image made by LFoundry of one SiPM pad that was marked as defective. The aluminum pad was etched in an uncontrolled way showing the layers at the bottom of the pad.}
\label{fig:SEMblackpad}
\end{figure}

Fig.~\ref{fig:hpareto_2} shows with a Pareto plot the sources of yield loss differentiating SiPMs according to their failure mechanism obtained by singularly applying the previously introduced requirement or by combining them. The largest source of yield loss is the Goodness of fit (G) cut, responsible for screening roughly 5\% of the tested SiPMs. The subsequent source of yield loss not yet presented in the previous section of this work, is labeled D in Fig.~\ref{fig:hpareto_2} and it is due to dice classified as out-of-spec at LFoundry's outgoing optical inspection because they have one or more aluminum pads etched in an uncontrolled way during processing. An example of a defective failure is shown in Fig.~\ref{fig:SEMblackpad}.
Note that these dice do not necessarily fail the characterization tests reported in previous sections of this work. Still, they may be problematic for other steps in the photosensor process flow ({\it e.g.} wire bonding). Rather than risk failures of entire tiles due to unforeseen issues with this defect, these dice are discarded, contributing a yield loss on the order of 1\% in the overall wafer production. \\Fig.~\ref{fig:hpareto_2} also shows other sources of yield losses ordered by their progressively lower efficiency in screening SiPMs. Note that the SiPMs discarded only due to their breakdown voltage or quenching resistor represent a negligible fraction. 

\begin{figure}[ht]
\centering
\includegraphics[width=0.99\linewidth]{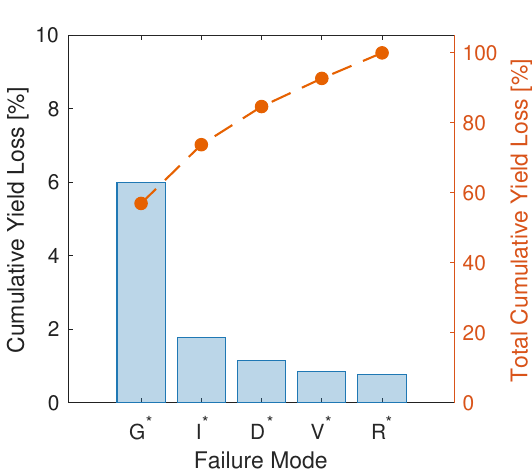}
\caption{Pareto plot that shows the cumulative source of the wafer yield loss. Each failure mechanism is treated as an independent quantity. The following yield loss mechanisms have been considered: (i) Goodness of fit ($\text{G}^{*}$), (ii) leakage current below breakdown ($\text{I}^{*}$), (iii) defective pads ($\text{D}^{*}$), (iv) quenching resistor ($\text{R}^{*}$), and (v) breakdown voltage ($\text{V}^{*}$).
}
\label{fig:hpareto}
\end{figure}

Fig.~\ref{fig:hpareto} shows again with a Pareto plot the same data of Fig.~\ref{fig:hpareto_2}, but each of the failure mechanisms is treated as an independent quantity showing for each of the selection criteria the cumulative yield loss due to each requirement {\it i.e.} SiPMs can fail for one or more requirements at the same time.

\section{Wafer Campaign Control Plans}

The DS-20k wafer testing campaign follows rigorous handling, storage and testing procedures. Wafer Lots are stored for their entire life in the NOA ISO-6 clean room (Sec.~\ref{sec:introduction}) sealed under vacuum anti-static bags. Storage parameters, such as humidity (HM), temperature (T), and radon contamination (RC), are monitored to ensure compliance with the DS-20k exposure specification ($\text{HM}<50\%,\text{T}<\SI{25}{C^\circ}, \text{RC}<\SI{50}{Bq/m^3}$). 

During wafer handling, only certified tools, such as SPS vacuum pencils, are used to touch the gold back surface of the wafers (Fig.~\ref{fig:wafercassette}), while the wafer front surface is touched only with specifically designed tools {\it e.g.} the frame mounter for dicing operation (Fig.~\ref{fig:wafer}).
The results of the wafer testing campaign are constantly monitored by using Statistical Process Control (SPC) Charts to detect trends or correlations with equipment parts ({\it e.g.} probe card needles degradation) and therefore to ensure the reliability of the cryo-probing operation over time. Moreover,  process capability indices are used to measure the ability of the cryoprobe process to produce SiPMs within the specification limits. 

More generally we monitor, on a weekly basis,
three process Capability (Cp) indexes linked to the specifications presented in the previous sections of this work: CpL, CpU and Cpk. CpL(U) is a measurement of the cryoprobing process based on its Lower (Upper) Specification Limit LSL(USL). CpL and CpU are defined in Eq.~\ref{eq:CPL} and Eq.~\ref{eq:CPU} as the ratio of two values: (i) the distance between the mean of the measured quantity $X$ and the lower (upper) specification limit LSL (USL), and (ii) the standard deviation of the process ($3\sigma_X$ variation)
\begin{equation}
\label{eq:CPL}
\text{CpL}=\frac{\overline{X}-\text{LSL}}{3\sigma_X}
\end{equation}

\begin{equation}
\label{eq:CPU}
\text{CpU}=\frac{\text{USL}-\overline{X}}{3\sigma_X}
\end{equation}

\begin{figure*}[ht]
\centering\includegraphics[width=0.9\linewidth]{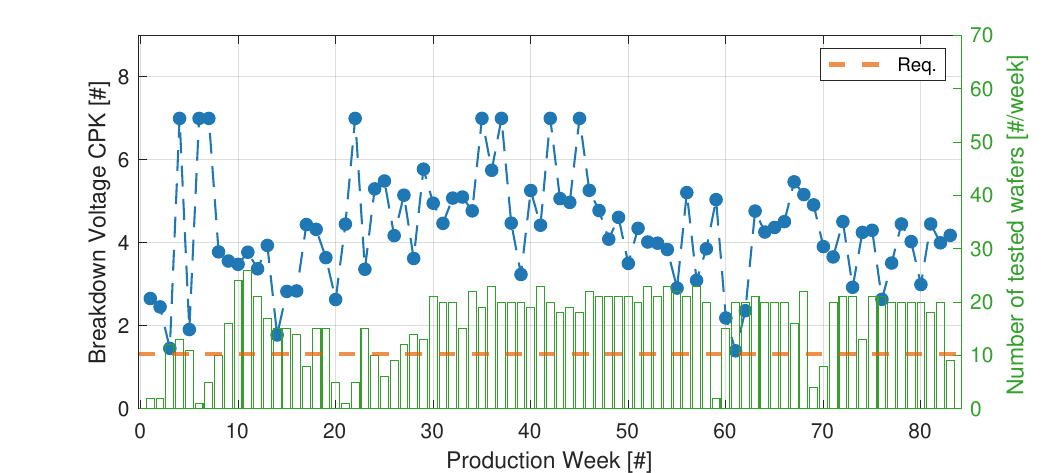}
\caption{Weekly based Cpk of the SiPM breakdown voltage (blue dots).  The dashed orange line represents the industry standard of 1.33. The green bars are the number of wafers tested in each production week.}
\label{fig:CPKVBD}
\end{figure*}

When upper and lower specifications are both needed to determine the process compliance, as is the case for breakdown voltage and quenching resistor specifications (Sec.~\ref{sec:expdetail}), the CpL(U) index is substituted by the Cpk, which is defined as the ratio between: (i) the difference between the mean of the measured quantity $X$ and the closest specification limit (USL or LSL), and (ii) the standard deviation of the process ($3\sigma_X$ variation):
\begin{equation}
\text{Cpk}=\text{min}\bigg[\frac{\text{USL}-\overline{X}}{3\sigma_X},\frac{\overline{X}-\text{LSL}}{3\sigma_X}\bigg]
\end{equation}
\begin{figure*}[ht]
\centering\includegraphics[width=0.9\linewidth]{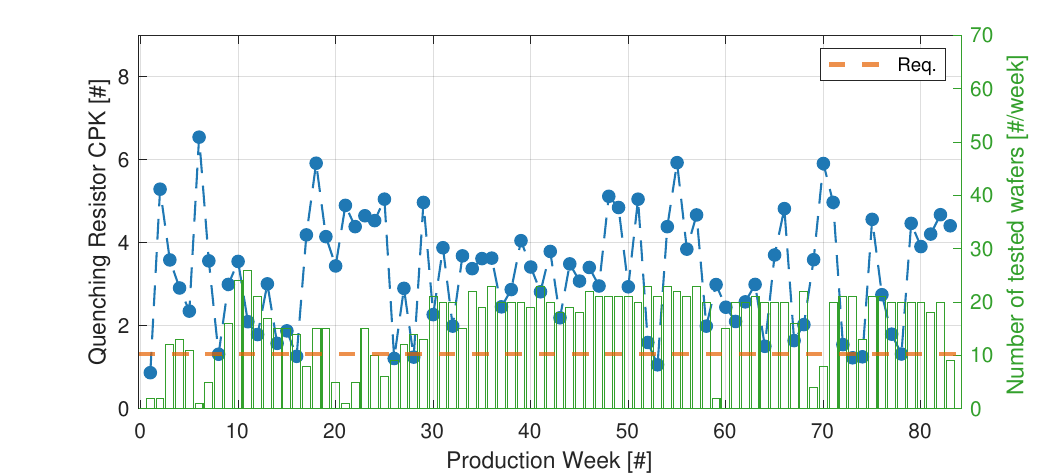}
\caption{Weekly Cpk of the SiPM quenching resistor (blue dots). The dashed orange line represents the industry standard of 1.33. The green bars are the number of wafers tested in each production week.}
\label{fig:CPKRQ}
\end{figure*}

Figs.~\ref{fig:CPKVBD} and \ref{fig:CPKRQ} display the weekly Cpk values for the SiPM breakdown voltage and quenching resistor. The horizontal line at 1.33 represents the industry standard requirement~\cite{842121}, which defines a $4\sigma_X$ difference between the mean of the measured quantity and its upper specification limit\footnote{Note that the Cpk is computed using the weekly median and standard deviation, with the standard deviation calculated after excluding outliers (defined as points more than 1.5 IQR above the upper quartile), as detailed in Sec.~\ref{sec:Vbd}.}. The figures also show the number of wafers tested each week. This number varies due to periodic maintenance of the cryoprobing system, which temporarily reduces weekly throughput. It is important to note that the early weeks of operation exhibited artificial variability because only about 20\% of the 25 wafers in each Lot were tested before moving on to the next Lot. This strategy was employed to efficiently sample the entire LFoundry wafer production and to assess the expected dominant Lot-to-Lot variability (Sec.~\ref{sec:expdetail}). Figs.~\ref{fig:CPKRQ} and \ref{fig:bRQ} further reveal a correlation with the Lot-to-Lot variation in the quenching resistor. Additionally, Fig.~\ref{fig:CPKVBD} and \ref{fig:CPKRQ} occasionally show exceptionally high Cpk values (greater than 5), which are linked to the testing procedure—for instance, during weeks when only a single wafer was tested (such as week 6 or 21). In contrast, single wafers or single Lots tend to exhibit much tighter distributions for breakdown voltage and quenching resistor, as demonstrated in Sec.~\ref{sec:Vbd} and Sec.~\ref{sec:Rq}. Overall, the process capability indices indicate a stable testing campaign, underscoring the ability of the cryoprobing process to consistently produce SiPMs within the specified limits.

\section{Conclusions}
This paper describes measurements performed in support of the DarkSide-20k experiment to characterize the properties of 1400 wafers of FBK NUV-HD-cryo (near-UV sensitive, high-density, cryogenic compatible) SiPMs produced by LFoundry. These wafers will be used to instruments the two $10.5~\text{m}^2$ DS-20k TPC optical planes and the inner and outer veto detectors. This work focused on the quality control and quality assurance of 1360 wafers (\NWAFER production, 46 engineering), representing \PERCENTPRODUCTION\% of the overall production. The data presented here encompass 56 of the 57 production lots, and include two engineering lots used for preproduction assembly, thereby capturing the Lot-to-Lot variation—the primary source of production variability. We measure an average breakdown voltage at 77~K of  $\overline{V}_\mathrm{bd}=27.19\pm0.05$~V, where the central value represents the distribution median and the error half of its InterQuartile Range, that is comfortably within the DS-20k specification ($27.2\pm1$~V). We measure an average quenching resistor value, leakage current below breakdown, and Goodness of Fit (GOF) parameter, all at 77~K, of $\overline{R}_\mathrm{q}=3.34\pm 0.15$~$\text{M}\upOmega$, $\overline{I}_L=6.6\pm2.2~\text{pA}$, and $\overline{\text{GOF}}=1.36\pm1.49$. These values are consistent with the DS-20k specifications ($I_L\le40~\text{pA}$,~$R_\mathrm{q}=3.35\pm1.50~\text{M}\upOmega$,~$\text{GOF}\le20$). In this work, we also show a correlation between the breakdown voltages and the quenching resistor values at room temperature and 77~K. More than 99\% of the dice that are within specification at room temperature are also compliant at cryogenic temperature. The low statistics of failing dice however do not allow us to conclude that the room temperature screening is sufficient to identify failing dice at cryogenic temperature. This is a consequence of the high wafer yield and low statistics of failing dice of the LFoundry production. Finally, we presented process capability indices that monitor the stability of the cryoprobing operation over time for all the tested wafer-level parameters.  The average production wafer yield is $93.2\pm2.5$\%, which exceeds the 80\% specification defined in the DarkSide-20k production plan. The wafer testing campaign is nearly complete, with fewer than 10\% of the wafers remaining to be tested. Following wafer-level testing, tile assembly and PDU assembly are progressing at NOA and in the UK, respectively. A dedicated QA/QC and characterization program is now in place, with the final step being a comprehensive cryogenic test of fully assembled units. These tests are being carried out at various collaboration institutions: the TPC PDUs are tested at INFN Naples, while the veto PDUs are evaluated in Liverpool, Edinburgh, Lancaster, and Warsaw.

\vspace{1cm}
\section{Acknowledgment}
\small{The authors thank E.~Citti, A.~Giannini, M.~Billaud (Keithley) and R.~Krippendorf, A.~Jachmann, R.~Liebschner, T.~Welisch, M.~Klos, D.~T. Knobbe (FormFactor), R.~Radice (ElectronMec) for the help during the commissioning of the S500 Keithley System and the PAC200 probe-station, respectively. The author also thank M.~Corosu (INFN Genova), C.~Pellegrino (INFN-CNAF) and the entire LNGS computing team (S.~Parlati, S.~Stalio, N.~Taborgna, R.~Giuliani, P.~Giuliani) for the support during the commissioning of the database and the network infrastructure at LNGS. The authors gratefully acknowledge support by the U. S. National Science Foundation (NSF) (Grants No. PHY-0919363, No. PHY-1004054, No. PHY-1004072, No. PHY-1242585, No. PHY-1314483, No. PHY-2310091, No. PHY- 1314507,
associated collaborative grants, No. PHY-1211308, No. PHY-1314501, No. PHY-1455351 and No. PHY-1606912, as well as Major Research
Instrumentation Grant No. MRI-1429544), the Italian Istituto Nazionale di Fisica Nucleare (Grants from Italian Ministero dell'Istruzione, Universita, e Ricerca Progetto Premiale 2013 and Commissione Scientific Nazionale II), the Natural Sciences and Engineering Research Council of Canada, SNOLAB, and the Arthur B. McDonald Canadian Astroparticle Physics Research Institute.  We acknowledge the financial support by LabEx UnivEarthS (ANR-10-LABX-0023 and ANR18-IDEX-0001), Chinese Academy of Sciences (113111KYSB20210030) and National Natural Science Foundation of China (12020101004). This work has been supported by the S\~{a}o Paulo Research Foundation (FAPESP) grant 2021/11489-7. I.~Albuquerque is
partially supported by Conselho Nacional de Desenvolvimento Cient\'{\i}fico e Tecnol\'ogico (CNPq). The authors were also supported by the Spanish Ministry of Science and Innovation (MICINN) through the grant PID2019-109374GB-I00, the Atraccion de Talento grant 2018-T2/TIC-10494, the Polish NCN (Grants No. UMO-2023/51/ B/ST2/02099, UMO-2023/50/A/ST2/00651 and 2022/47/B/ST2/02015), the Polish Ministry of Science and Higher Education (MNiSW, grant number 6811/IA/SP/2018), the International Research Agenda Programme AstroCeNT (Grant No. MAB/2018/7) funded by the Foundation for Polish Science from the European Regional Development Fund, the European Union Horizon 2020 research and innovation program under grant agreement No 952480 (DarkWave), the Science and Technology Facilities Council, part of the United Kingdom Research and Innovation, The Royal Society (United Kingdom), and IN2P3-COPIN consortium (Grant No. 20-152). We also wish to acknowledge the support from Pacific Northwest National Laboratory, which is operated by Battelle for the U.S. Department of Energy under Contract No. DE-AC05-76RL01830. This research was also supported by the Fermi National Accelerator Laboratory (Fermilab), a U.S. Department of Energy, Office of Science, HEP User Facility. Fermilab is managed by Fermi Research Alliance, LLC (FRA), actstg under Contract No. DE-AC02-07CH11359.}

\bibliographystyle{unsrt}
\bibliography{bibliography}

\newcommand{\notds}{\nolinebreak\footnotemark\nolinebreak}
\renewcommand{\thefootnote}{$*$}
{
\onecolumn
\textbf{The DarkSide-20k Collaboration}

F.~Acerbi\thanksref{l_TNFBK}\nolinebreak,
P.~Adhikari\thanksref{l_Carleton}\nolinebreak,
P.~Agnes\thanksref{l_AQGSSI}\textsuperscript{,}\thanksref{l_AQLNGS}\nolinebreak,
I.~Ahmad\thanksref{l_AstroCeNT}\nolinebreak,
S.~Albergo\thanksref{l_CTINFN}\textsuperscript{,}\thanksref{l_CTUNI}\nolinebreak,
I.~F.~Albuquerque\thanksref{l_USP}\nolinebreak,
T.~Alexander\thanksref{l_PNNL}\nolinebreak,
A.~K.~Alton\thanksref{l_Augustana}\nolinebreak,
P.~Amaudruz\thanksref{l_TRIUMF}\nolinebreak,
M.~Angiolilli\thanksref{l_AQGSSI}\textsuperscript{,}\thanksref{l_AQLNGS}\nolinebreak
% ...

\begin{enumerate}
\item{\TNFBK\label{l_TNFBK}}
\item{\Carleton\label{l_Carleton}}
\item{\label{l_AQGSSI}\AQGSSI}
\item{\label{l_AQLNGS}\AQLNGS}
\item{\AstroCeNT\label{l_AstroCeNT}}
\item{\CTINFN\label{l_CTINFN}}
\item{\CTUNI\label{l_CTUNI}}
\item{\USP\label{l_USP}}
\item{\PNNL\label{l_PNNL}}
\item{\Augustana\label{l_Augustana}}
\item{\TRIUMF\label{l_TRIUMF}}
\item{\Columbia\label{l_Columbia}}
\item{\CAINFN\label{l_CAINFN}}
\item{\CAUniPHY\label{l_CAUniPHY}}
\item{\Alberta\label{l_Alberta}}
\item{\MendeleevUniverisity\label{l_MendeleevUniverisity}}
\item{\LNLINFN\label{l_LNLINFN}}
\item{\SNL\label{l_SNL}}
\item{\RHUL\label{l_RHUL}}
\item{\CIEMAT\label{l_CIEMAT}}
\item{\CPPM\label{l_CPPM}}
\item{\PIINFN\label{l_PIINFN}}
\item{\PIUniPHY\label{l_PIUniPHY}}
\item{\Oxford\label{l_Oxford}}
\item{\TOINFN\label{l_TOINFN} }
\item{\TOPoli\label{l_TOPoli}}
\item{\RMUnoINFN\label{l_RMUnoINFN}}
\item{\GEUni\label{l_GEUni}}
\item{\GEINFN\label{l_GEINFN}}
\item{\MIPoliICA\label{l_MIPoliICA}}
\item{\MIINFN\label{l_MIINFN}}
\item{\WUT\label{l_WUT} }
\item{\RMTreINFN\label{l_RMTreINFN} }
\item{\NAINFN\label{l_NAINFN} }
\item{\VTech\label{l_VTech} }
\item{\CAUniEEE\label{l_CAUniEEE} }
\item{\Zaragoza\label{l_Zaragoza} }
\item{\MSU\label{l_MSU} }
\item{\BOINFN\label{l_BOINFN} }
\item{\BOUniPHY\label{l_BOUniPHY} }
\item{\Laurentian\label{l_Laurentian} }
\item{\SNOLAB\label{l_SNOLAB} }
\item{\UnivAQ\label{l_UnivAQ} }
\item{\Krakow\label{l_Krakow} }
\item{\NAUniDIST\label{l_NAUniDIST} }
\item{\MIUni\label{l_MIUni} }
\item{\RMUnoUni\label{l_RMUnoUni} }
\item{\MIPoliCHE\label{l_MIPoliCHE} }
\item{\Petersburg\label{l_Petersburg} }
\item{\NAUniPHY\label{l_NAUniPHY} }
\item{\Queens\label{l_Queens} }
\item{\Princeton\label{l_Princeton} }
\item{\Kurchatov\label{l_Kurchatov} }
\item{\Belgorod\label{l_Belgorod} }
\item{\UCDavis\label{l_UCDavis} }
\item{\Lancaster\label{l_Lancaster} }
\item{\APC\label{l_APC} }
\item{\Liverpool\label{l_Liverpool} }
\item{\BINP\label{l_BINP} }
\item{\BOCentroFermi\label{l_BOCentroFermi} }
\item{\Manchester\label{l_Manchester} }
\item{\ETHZ\label{l_ETHZ} }
\item{\WilliamsCollege\label{l_WilliamsCollege} }
\item{\Hawaii\label{l_Hawaii} }
\item{\Chicago\label{l_Chicago} }
\item{\CTLNS\label{l_CTLNS} }
\item{\IHEP\label{l_IHEP} }
\item{\UniversityofEdinburgh\label{l_UniversityofEdinburgh} }
\item{\Houston\label{l_Houston} }
\item{\NAUniPHARM\label{l_NAUniPHARM} }
\item{\UCRiverside\label{l_UCRiverside} }
\item{\JINR\label{l_JINR} }
\item{\MEPhI\label{l_MEPhI} }
\item{\Birmingham\label{l_Birmingham} }
\item{\UniHAM\label{l_UniHAM} }
\item{\LFoundry\label{l_LFoundry} }
\item{\Temple\label{l_Temple} }
\item{\Warwick\label{l_Warwick} }
\item{\INFN\label{l_INFN} }
\item{\UMass\label{l_UMass} }
\item{\NAUniCHE\label{l_NAUniCHE} }
\item{\CAUniCHE\label{l_CAUniCHE} }
\item{\UCAS\label{l_UCAS} }
\item{\UCLA\label{l_UCLA} }
\item{\FortLewis\label{l_FortLewis} }
\item{\Lodz\label{l_Lodz}}
\item{\STFCInterconnect\label{l_STFCInterconnect}}
\item{\ENUniCEE\label{l_ENUniCEE}}
\item{\Seattle\label{l_Seattle}}
\end{enumerate}

}

\end{document}